\documentclass[aps,prd,12pt]{revtex4}

\begin{document}
\def\ER{E_{\rm R}}
\def\eth{\epsilon}
\def\emx{\varepsilon}
\def\SD{{\rm SD}}
\def\SI{{\rm SI}}
\renewcommand{\baselinestretch}{1.0}

\title{Nuclear spin structure in dark matter search: \protect\\
       The finite momentum transfer limit} 

\author{V.A.~Bednyakov}
\affiliation{Dzhelepov Laboratory of Nuclear Problems,
         Joint Institute for Nuclear Research, \\
         141980 Dubna, Russia; E-mail: Vadim.Bednyakov@jinr.ru}
\author{and F. \v Simkovic}
\affiliation{Department of Nuclear Physics, Comenius University,
Mlynsk\'a dolina F1, SK--842 15 Bratislava, Slovakia}

\date{\today}

\begin{abstract}
        Spin-dependent elastic scattering of weakly interacting 
	massive dark matter particles (WIMP) off nuclei is reviewed.
	All available, within different nuclear models, 
	structure functions $S(q)$ 
	for {\em finite}\/ momentum transfer ($q>0$) 
	are presented. 
	These functions describe the recoil energy dependence of the 
	differential event rate 
	due to the spin-dependent WIMP-nucleon interactions.
	This paper, together with the previous paper 
	``Nuclear spin structure in dark matter search: 
        The zero momentum transfer limit'',
	completes our review of 
	the nuclear spin structure calculations involved 
	in the problem of direct dark matter search.

 

\end{abstract}

\maketitle

\section{Introduction}
    Weakly Interacting Massive Particles (WIMPs) are among the
    most popular candidates for the relic cold dark matter (DM).
    There is some revival of interest in the 
    WIMP-nucleus spin-dependent interaction from both theoretical 
    (see e.g. 
\cite{Engel:1991wq,Bottino:2003cz,%
Bednyakov:1994te,Bednyakov:2000he,Bednyakov:2002mb,Bednyakov:2003wf,%
Bednyakov:2004xq,Bednyakov:2005qp})
     and experimental (see e.g. 
\cite{Girard:2005pt,Girard:2005dq,%
Giuliani:2004uk,Giuliani:2005bd,%
Savage:2004fn,Benoit:2004tt,Tanimori:2003xs,Ovchinnikov:2003AA,%
Moulin:2005sx,Mayet:2002ke,Klapdor-Kleingrothaus:2005rn})
      points of view.  
     There are some proposals aimed at direct DM detection with 
     relatively low-mass isotope targets 
\cite{Girard:2005pt,Girard:2005dq,Tanimori:2003xs,Ovchinnikov:2003AA,%
Moulin:2005sx,Mayet:2002ke}
      as well as some attempts to design and construct 
      a DM detector which is sensitive to the nuclear recoil direction
\cite{Alner:2004cw,Snowden-Ifft:1999hz,Gaitskell:1996cv,Sekiya:2004ma,%
Morgan:2004ys,Vergados:2000cp,Vergados:2002bb}.
      Low-mass targets 
      make preference for the low-mass WIMPs 
      and are more sensitive to the spin-dependent 
      WIMP-nucleus interaction as well 
\cite{Jungman:1996df,Engel:1991wq,Divari:2000dc,%
Bednyakov:1994te,Bednyakov:2000he,Bednyakov:2004xq,Bednyakov:1997ax}.
        On the other hand, WIMPs with masses about 100 GeV/$c^2$ 
	follow from the results of the DArk MAtter (DAMA) experiment.
	This collaboration claimed 
        observation of the first evidence for the dark matter signal
        due to registration of the annual modulation effect
\cite{Bernabei:2000qi,Bernabei:2003za,Bernabei:2003wy}.
	Aimed for more than one decade at 
	the DM particle direct detection, the DAMA experiment 
	with 100-kg highly radio-pure NaI(Tl) scintillator detectors  
	successfully operated till July 2002 
	at the Gran Sasso National Laboratory of the I.N.F.N.
	On the basis of the results obtained for over 7 annual cycles 
	(107731~kg$\cdot$day total exposure)
	the effectiveness of the WIMP model-independent 
	annual modulation signature was demonstrated 
	and the WIMP presence in the galactic halo is strongly supported 
	at 6.3 $\sigma$ C.L. 
\cite{Bernabei:2003za}.

	The goal of this review 
	(being a continuation of our previous review paper
\cite{Bednyakov:2004xq})
        is to complete our nuclear physics consideration of   
	the spin-dependent (or axial-vector) interaction of 
	dark matter particles with nuclei. 
	This type of interaction of the DM particles is important for the 
	following reasons:
	(i) the spin-dependent interaction  of the DM 
        particles provides us with twice stronger constraints on the SUSY 
        parameter space in comparison with the spin-independent interaction; 
        (ii) in the case of spin-dependent interaction of heavy WIMPs
        with heavy target nuclei 
	the so-called long $q$-tail behavior of the relevant form--factor 
        allows detection of large nuclear recoil energy due to some
        nuclear structure effects;
        (iii) it is worthwhile to note that by relying only upon
        the scalar interaction of the DM particles, which seems to
        be strongly suppressed, one might miss a DM signal
\cite{Bednyakov:2000he}. However, by a simultaneous study of both 
        spin-dependent and spin-independent interactions of the
        DM particles with nuclei the chance for observing the
        DM signal is significantly increased.

        A dark matter event is elastic scattering of a relic WIMP 
	(or neutralino) $\chi$ 
	with mass ${m_\chi}$ from a target nucleus $A$ producing a nuclear 
	recoil $E_{\rm R}$ which can be detected by a suitable detector.
	The differential event rate in respect to the recoil 
	energy is the subject of experimental measurements.
	The rate depends on the distribution of
        the relic WIMPs 
	in the solar vicinity $f(v)$ and
        the cross section of WIMP-nucleus elastic scattering
\cite{Jungman:1996df,Lewin:1996rx,Smith:1990kw,Bednyakov:1999yr,Bednyakov:1996yt,Bednyakov:1997ax,Bednyakov:1997jr,Bednyakov:1994qa}.
	The differential event rate per unit mass of 
	the target material has the form
\begin{equation}
\label{Definitions.diff.rate}
	\frac{dR}{dE_{\rm R}} = N_T \frac{\rho_\chi}{m_\chi}
	\int^{v_{\max}}_{v_{\min}} dv f(v) v
	{\frac{d\sigma^A}{dq^2}} (v, q^2). 
\end{equation}
        We assume WIMPs (neutralinos) to be a dominant component of
        the DM halo of our galaxy with a density
        $\rho_{\chi}$ = 0.3 GeV$/$cm$^{3}$ in the solar vicinity.
        The nuclear recoil energy
	$E_{\rm R} = q^2 /(2 M_A )$ is typically about $10^{-6} m_{\chi}$ and 
	$N_T$ is the number density of a target nuclei with mass $M_A$. 
	The WIMP-nucleus differential 
	elastic scattering cross section 
	for spin-non-zero ($J\neq 0$) 
	nuclei contains coherent (spin-independent, or
	SI) and axial (spin-dependent, or SD) terms
\cite{Engel:1992bf,Engel:1991wq,Ressell:1993qm}: 
\begin{eqnarray} \label{drateEPV}
\frac{d\sigma^A}{dq^2}(v,q^2)
 &=&\frac{d\sigma^{A}_\SD}{dq^2}(v,q^2)
   +\frac{d\sigma^{A}_\SI}{dq^2}(v,q^2) 
  =  \frac{S^{A}_\SD (q^2)}{v^2(2J+1)} 
   +\frac{\sigma^A_\SI(0)}{4\mu_A^2 v^2}F^2_\SI(q^2).
\end{eqnarray}
	Here $\displaystyle \mu_A = \frac{m_\chi M_A}{m_\chi+ M_A}$
	is the reduced WIMP-nucleus mass, 
	$\sigma^A_\SI(0)=\frac{\mu_A^2}{\pi}[A^2 C_{0}^2]$ 
	is the spin-independent WIMP-nucleus total elastic cross section 
	at $q^2=0$ and $F^2_\SI(q^2)$ is the 
	normalized ($F^2_{\SI}(0) = 1$)
	nozero-momentum-transfer spin-independent nuclear form-factor.
	The $q^2$-dependence of the SD cross section is 
	governed by the spin-dependent 
	structure function $S^A_\SD(q)$ of Engel et al.
\cite{Engel:1992bf,Engel:1991wq}
\begin{eqnarray}
\label{SF-definition}
S^A_\SD(q) = S^A(q)
&=& \sum_{L\ \rm odd} \big( \vert\langle N \vert\vert {\cal T}^{el5}_L
(q) \vert\vert N \rangle\vert^2 + \vert\langle N \vert\vert {\cal L}^5_L
(q) \vert\vert N \rangle\vert^2\big).
\end{eqnarray}
	The transverse electric ${\cal T}^{el5}(q)$ 
	and longitudinal ${\cal L}^5(q)$ multipole projections of the
	axial vector current operator are given in the Appendix.

	Relic WIMPs in the halo of our Galaxy have 
	a mean velocity of
	$\langle v \rangle  \simeq 300~{\rm km/s} = 10^{-3} c$.  
	When $R\ll 1/q^{}_{\max}$ 
	(or $q_{\max}R \ll 1$)
	where $R$ is the nuclear size and 
	$q_{\max} = 2 \mu_A v$ is the maximum 
	of the momentum transfer in the  
	{$\chi$}-nucleus scattering, 
	the spin structure function $S^A(q)$ reduces to
(so-called {\em zero transfer momentum limit})
\begin{eqnarray*}
S^A(0)={1\over{4 \pi}} \vert\langle A \vert\vert\sum_i
    {1\over{2}}(a_0 + a_1 \tau_3^i) {\bf \sigma}_i\vert\vert A \rangle\vert^2 
     =\frac{2J+1}{\pi} J(J+1) \Lambda^2 
\quad {\rm with} \quad 
\Lambda = {{a_p \langle {\bf S}_p \rangle}\over{J}}  +
{{a_n \langle {\bf S}_n \rangle}\over{J}}.
\end{eqnarray*}
	Here $\langle {\bf S}_{p,n} \rangle$ is the proton (or neutron) spin
	averaged over nucleus $A$, 
	$a_n$ and $a_p$ are the
	effective spin WIMP-neutron and WIMP-proton couplings
	that contain details of the supersymmetric model, as
	well as the quark spin content of the proton and neutron.

	As $m_{\chi}$ increases, $R\approx 1/q$
	(the product $qR$ starts to become non-negligible) 
	and 
	{\em the finite momentum transfer limit}\/
	must be considered for heavier nuclei.  
	The formalism is a straight forward extension of that
	developed for the study of weak and electromagnetic 
	semi-leptonic interactions in nuclei. 
	Here we follow the definitions of 
\cite{Ressell:1997kx,Ressell:1993qm}.
	With the isoscalar spin coupling constant $a_0 = a_n + a_p$
	and the isovector spin coupling constant
	$a_1 = a_p - a_n$ one can split 
	the nuclear structure function $S^A_{}(q)$ into a pure
	isoscalar term, $S^A_{00}(q)$, a pure isovector term, $S^A_{11}(q)$, 
	and an interference term, $S^A_{01}(q)$, in the following way:
\begin{equation}
\label{Definitions.spin.decomposition}
S^A_{}(q) = a_0^2 S^A_{00}(q) + a_1^2 S^A_{11}(q) + a_0 a_1 S^A_{01}(q).
\end{equation}
	These three partial structure functions
	contain expectation values of operators of the form
	$j_L(qr) [ Y_L \sigma]^{L \pm 1}$, 
	which depend on spatial coordinates and nucleon spins.
	The relations 
\begin{eqnarray}
\label{SF-normalization}
S^A_{00}(0) &=& C(J)(\langle {\bf S}_p \rangle + \langle {\bf S}_n \rangle)^2,
\quad
S^A_{11}(0) = C(J)(\langle {\bf S}_p \rangle - \langle {\bf S}_n \rangle)^2,
\\ \nonumber
S^A_{01}(0) &=&2C(J)(\langle {\bf S}^2_p \rangle - \langle {\bf S}^2_n \rangle)
\quad
{\rm with}\quad 
C(J)= \frac{2J+1}{4\pi}\frac{J+1}{J},
\end{eqnarray}
       connect the nuclear spin structure function $S^A(q)$ at $q=0$ 
       with proton $\langle {\bf S}_p \rangle$ 
       and neutron $\langle {\bf S}_n \rangle$
       spin contributions averaged over the nucleus. 
       In relations
(\ref{SF-normalization}) the normalization coefficient
       $C(J)>0$ and therefore $S^{}_{00}(0)\ge 0$ and $S^{}_{11}(0)\ge 0$.
       These three partial structure  functions $S^A_{ij}(q)$ 
       allow calculation 
       of spin-dependent cross sections for any heavy Majorana particle
       as well as for the neutralino with arbitrary composition
\cite{Engel:1995gw}.

\smallskip
        In this paper we extend our consideration 
	of the modern nuclear spin structure calculations involved 
	into the problem of the direct dark matter search. 
	The calculations of the proton and neutron spins
	$ \langle {\bf S}_{p(n)} \rangle $ 
	averaged over all nucleons in the nucleus $A$,
	which are relevant to the zero-momentum 
	neutralino-nucleon spin cross sections,
	are considered in our previous review  
\cite{Bednyakov:2004xq}.
	The cross sections at zero momentum transfer show strong 
	dependence on the nuclear structure of the ground state
\cite{Divari:2000dc}. 
	Here we discuss the calculations of the spin structure
	functions in the {\em finite}\/ momentum transfer approximation.
	We also touch upon the 
	level of accuracy of these calculations.
        Finally, (in the last section) 
	we briefly discuss a new approach to 
        data analysis in the finite momentum transfer approximation 
	directly in terms of the effective spin nucleon couplings 
	$a_{0,1}$ together with the scalar WIMP-proton cross
	section $\sigma^{p}_\SI(0)$ at $q=0$.
       Contrary to other possibilities, 
       (see for example, 
\cite{Tovey:2000mm}), 
       this procedure is direct and relies 
       as much as possible on the results of the 
       most accurate nuclear spin structure calculations.
\bigskip

\section{Non-zero momentum transfer limit}
	To the best of our knowledge,  
        the finite, non-zero momentum transfer 
	calculations of the spin nuclear structure functions $S^A_{}(q)$
	have been performed for the set of isotopes given in 
Table~\ref{List.SD-non-0}.
\begin{table}[h!]
\caption{List of isotopes with available 
spin structure functions, $S^A_{}(q)$, at $q>0$.}
\label{List.SD-non-0} 
\bigskip
\begin{tabular}{|r|l|l|} \hline
 A  &~~~~~Isotope  &~~~~Authors and reference(s) \\ \hline
19  & Fluorine, $^{19}$F  
    & Vergados et al. \cite{Vergados:2002bb,Divari:2000dc,Vergados:1996hs} \\
\hline
23  & Sodium, $^{23}$Na 
    & Ressell and Dean \cite{Ressell:1997kx} \\ 
   && Vergados et al. \cite{Ressell:1997kx,Divari:2000dc}\\
\hline
27  & Aluminium, $^{27}$Al 
    & Engel et al. \cite{Engel:1995gw} \\ 
\hline
29  & Silicon, $^{29}$Si
    & Ressell et al. \cite{Ressell:1993qm} \\ 
    && Vergados et al. \cite{Vergados:2002bb,Divari:2000dc} \\
\hline
39  & Potassium, $^{39}$K 
    & Engel et al. \cite{Engel:1995gw} \\
\hline
73  & Germanium, $^{73}$Ge~~~~
    & Ressell et al. \cite{Ressell:1993qm} \\ 
    && Demitrov et al. \cite{Dimitrov:1995gc}\\ 
\hline
93  & Niobium, $^{93}$Nd 
    & Engel et al. \cite{Engel:1992qb} \\
\hline
123 & Tellurium, $^{123}$Te 
    & Nikolaev and Klapdor-Kleingrothaus \cite{Nikolaev:1993vw}~~~~~\\
\hline
125 & Tellurium, $^{125}$Te
    & Ressell and Dean \cite{Ressell:1997kx}\\ 
\hline
127 & Iodide, $^{127}$I
    & Ressell and Dean \cite{Ressell:1997kx}\\ 
\hline
129 & Xenon, $^{129}$Xe 
    & Ressell and Dean \cite{Ressell:1997kx}\\ 
\hline
131 & Xenon, $^{131}$Xe  
    & Engel \cite{Engel:1991wq} \\ 
    && Ressell and Dean \cite{Ressell:1997kx} \\ 
    && Nikolaev and Klapdor-Kleingrothaus \cite{Nikolaev:1993vw} \\ 
\hline
~207&Lead, $^{207}$Pb 
    & Vergados and Kosmos \cite{Kosmas:1997jm,Vergados:1996hs} \\
\hline
\end{tabular}
\end{table}
	The zero-momentum transfer limit (mostly quenching) 
	is also investigated for Cd, Cs, Ba and La 
\cite{Pacheco:1989jz,Nikolaev:1993vw,Iachello:1991ut},
        for hydrogen, $^1$H,
\cite{Ellis:1988sh,Ellis:1991ef}, 
	helium, $^3$He, 
\cite{Vergados:1996hs},  
	chlorine, $^{35}$Cl,
\cite{Ressell:1993qm}
        and calcium, $^{43}$Ca,
\cite{Lewin:1996rx}. 

       General discussion of the nuclear model approaches to 
       calculation of the spin characteristics like spins
       averaged over the nucleus proton (neutron) 
       $ \langle {\bf S}_{p(n)} \rangle $, 
       one can find in our previous paper
\cite{Bednyakov:2004xq}.
       In this paper all available
       sets of the spin structure functions are given 
       either in the form of explicit functions 
       or as useful analytical parameterizations of 
       the accurate numerical results, or only 
       graphically (as pictures from original papers).

        As already noted in the introduction 
	for quite heavy WIMPs 
	and sufficiently heavy nuclei, the dependence
	of the nuclear matrix elements on the momentum transfer cannot
	be ignored even if the WIMP has energies as low as 100~keV. 
	For example, 
	if $m_{\chi} \gg m_A$, 
	the reduced mass $\mu_A$ almost reaches $m_A$  
	($\mu_A \rightarrow m_A$).
	It is rather popular (and simplest) assumption that the WIMPs have a 
	Maxwellian velocity distribution in the halo of our Galaxy. 
	Some WIMPs will possess velocities significantly greater
	than $\langle v \rangle \simeq 10^{-3}c$.
	A maximum velocity
	of $v_{\max} \simeq 700$ km/s (slightly greater than the
 	Galactic escape velocity and more than twice the mean WIMP velocity)
	implies maximum
	momentum transfers of $q_{\max} \simeq 550$~MeV
	for nuclei with atomic weight $A \sim 127$. 
	This $q$ value is not {\it small}\/
	compared to the inverse nuclear size
\cite{Ressell:1997kx} and one has to use 
	{\em the finite momentum transfer approximation}\/
	for heavier nuclei.  

	Despite the above-mentioned simple kinematic estimation of $q_{\max}$ 
	(which is used over the text for illustration) 
	it is worth noting that this $q_{\max}$ value is in fact 
	too large and is almost not reachable.
	The WIMP-nucleus interaction is very weak, 
	it occurs very rarely and therefore
	the impulse approximation for rather large $q$
	is well motivated and 
	is used almost 
	in all nuclear calculations reviewed in the paper. 
	In the impulse approximation the WIMP-nuclear interaction 
	is described by means of the WIMP interaction with a 
	nucleon from the nucleus $A$ (see Appendix) 
	and the maximal momentum transfer 
	in the WIMP-nucleon system is 
	considerably smaller than the $q_{\max}$. 
	Therefore for heavy enough nuclei, in general, 
	the transfer momentum $q\ll q_{\max}$. 

	The full momentum dependence of the form factors must be calculated
	from detailed nuclear models, and the results 
	are especially important for heavier nuclei
\cite{Jungman:1996df}.
	Unfortunately, the simple phenomenological analysis 
	in the OGM (odd group model) and EOGM (extended OGM) of 
\cite{Engel:1989ix} cannot be extended to the finite momentum transfer case,
	because the experimental data directly related to 
	the neutralino-nucleus elastic scattering is not available 
\cite{Engel:1991wq}.
	Quite a number of high multipoles can now contribute, 
	some of them getting contributions from components of
	the wave function which do not contribute in the static limit
	(i.e. at $q=0$). 
	Thus, in general, sophisticated Shell Model calculations 
	are needed to account both for the observed retardation
	of the static spin matrix element and its correct dependence
	on transfer momentum, $q$. 
	For the experimentally interesting nuclear systems 
	$^{29}_{14}$Si and $^{73}_{32}$Ge 
	very elaborate calculations have been performed  by Ressell et al. 
\cite{Ressell:1993qm}. 
	In the case of $^{73}_{32}$Ge 
	a further improved calculation by 
	Dimitrov, Engel and Pittel was carried out 
\cite{Dimitrov:1995gc}
	by suitably mixing variationally determined 
	triaxial Slatter determinants. 
	Indeed, for this complex nucleus many multipoles 
	contribute and the needed calculations involve techniques which 
	are extremely sophisticated
\cite{Kosmas:1997jm}.
	Now the necessity for more detailed calculations {\em especially}\/ 
	for the spin-dependent component of the cross sections 
	for heavy nuclei is well motivated.

        Further available sets of spin
	structure functions $S^A(q)$ are collected
	for nuclei from $^{19}$F up to the $^{207}$Pb.

\subsection{Fluorine, $^{19}$F}
        Fluorine-19 is the isotope most sensitive to the spin-dependent
	WIMP-nuclear interaction 
\cite{Divari:2000dc} and a lot of experimental groups
        hope to explore this feature of $^{19}$F experimentally
(see, for example, 
\cite{Ogawa:2000vi,Boukhira:2002qj,Girard:2005dq,Tanimori:2003xs,%
Takeda:2003km,Miuchi:2002zp}).

        The finite momentum transfer spin structure function 
	$S^{19}(q)$ for $^{19}$F,
	one of the lightest DM target medium, 
	(together with other light targets $^{23}$Na and $^{29}$Si) 
	was obtained for the first time by Vergados with co-authors 
\cite{Divari:2000dc}.
	The spin contribution to the differential cross section was
        carefully estimated by the shell-model calculations in the $sd$ shell 
        using the Wildenthal interaction
(see, for example,
\cite{Brown:1985aa,Brown:1988aa}),
	which was developed and tested over many years. 
	This interaction is known to reproduce accurately 
	many nuclear observables for $sd$ shell nuclei. 
	The Wildenthal two-body matrix elements as well 
	as the single-particle energies are determined by 
	fits to experimental data in nuclei from $A=17$ to $A=39$.
        The shell-model wave functions used by the authors were tested 
        in the calculation of the low-energy spectra and 
	ground state magnetic moment.
	Rather good agreement between the theoretical and experimental  
	results was achieved
(see Tables 2--4 in our previous review
\cite{Bednyakov:2004xq}). 
        Their spin matrix elements 
        are in good agreement with those of previous calculations 
\cite{Ressell:1993qm}. 
 
        The pure isoscalar, $S^{19}_{00}$, pure isovector, $S^{19}_{11}$, 
	and interference, $S^{19}_{01}$ terms 
(\ref{Definitions.spin.decomposition}) of the 
        fluorine ($J=1/2$) 
	structure function $S^{19}(q)$ can be given in the form
\cite{Divari:2000dc}: 
\begin{eqnarray}
\nonumber
S^{19}_{00}(q) &=& \frac{2J+1}{16\pi} \times (2.610)\times 
          e^{-u}\left\{P^2_{(0,1)}(u) + P^2_{(2,1)}(u) \right\} , 
\\ \label{SF-F19}
S^{19}_{11}(q) &=& \frac{2J+1}{16\pi} \times (2.807) \times 
          e^{-u}\left\{ Q^2_{(0,1)}(u) + Q^2_{(2,1)}(u) \right\} , 
\\ \nonumber
S^{19}_{01}(q) &=& \frac{2J+1}{8\pi}\times (2.707)\times 
   e^{-u}\left\{P_{(0,1)}(u) Q_{(0,1)}(u) + P_{(2,1)}(u) Q_{(2,1)}(u) \right\}.
\end{eqnarray}
Here the following functions are introduced: 
$$
P_{(0,1)}(u) = 0.1145 u^2 - 0.6667 u + 1, \qquad 
Q_{(0,1)}(u)  = 0.1088 u^2 - 0.6667 u + 1,
$$ $$
P_{(2,1)}(u) = -0.0026\, u^2 + 0.0100\, u, \qquad  
Q_{(2,1)}(u) =  0.0006\, u^2 + 0.0041\, u.         
$$
       The dimensionless variable $u=(qb)^2/2$ and 
       the energy transfer to the nucleus $A$ are connected by 
       the relation $E = u b^{-2}/ M_A$ (or $u = b^{2}\,E\, M_A$).
       Here 
       $b=1.00A^{1/6} = 1.63\,$fm~$=1.63/0.1973$~GeV$^{-1}~$ 
       is the oscillator size parameter for $^{19}$F. 
       For $v_{\max} \approx 600$~km/s one has 
       $u_{\max} \approx 0.17$ (or $q_{\max}\approx 71$~MeV, and 
       $E_{\max} \approx 140$~keV) and  
       $u_{\min} \approx 0.011$, 
       which corresponds to the energy threshold of 10~keV 
       (and $q_{\min} \approx 17~$MeV).

       Following 
\cite{Divari:2000dc} we note 
       that at a relatively large momentum transfer the 
       nucleonic axial current gets modification (PCAC)
$\displaystyle {\vec\sigma} \to {\vec\sigma} - 
       \frac{(\vec\sigma \cdot \vec p)\vec p}{q^2+m^2_\pi} $. 
       Therefore structure functions 
(\ref{SF-F19}) become more complicated:
\begin{eqnarray}
\nonumber
S^{19}_{00}(q) 
&=& \frac{2.610}{8\pi}\, 
         e^{-u}\left\{P^2_{(0,1)}(u) (1+\beta_0(u)) 
                    + P^2_{(2,1)}(u) (1+\beta_2(u)) 
- 2 \beta_{02}(u) P_{(0,1)}(u) P_{(2,1)}(u) \right\} , 
\\ \nonumber
S^{19}_{11}(q) 
&=& \frac{2.807}{8\pi}\,
              e^{-u}\left\{ Q^2_{(0,1)}(u) (1+\beta_0(u)) 
                          + Q^2_{(2,1)}(u) (1+\beta_2(u))
- 2 \beta_{02}(u) Q_{(0,1)}(u) Q_{(2,1)}(u)\right\} , 
\\ 
\label{SF-F19-pi}
S^{19}_{01}(q) 
&=& \frac{2.707}{4\pi}\,
            e^{-u}\left\{
             P_{(0,1)}(u) Q_{(0,1)}(u) (1+\beta_0(u)) 
           + P_{(2,1)}(u) Q_{(2,1)}(u) (1+\beta_2(u)) \right. \\
\nonumber
   && \qquad\qquad \qquad\qquad 
      \left. -\beta_{02}(u)\times (
               P_{(0,1)}(u) Q_{(2,1)}(u)+Q_{(0,1)}(u) P_{(2,1)}(u) )
\right\}, 
\end{eqnarray}
     where 
\begin{equation}
\label{Pion-finite-mass}
\beta_0(u)=\frac13 \left[ \left(\frac{u_\pi}{u+u_\pi}\right)^2-1\right], 
\quad
\beta_2(u)= 2\beta_0(u), 
\quad
\beta_{02}(u)= - \sqrt{2}\beta_0(u), 
\quad u_\pi=\frac{(bm_\pi)^2}{2}. 
\end{equation}
\begin{figure}[h!] 
\begin{minipage}[b]{0.45\textwidth}
{\begin{picture}(50,75)
\put(-50,-87){\includegraphics{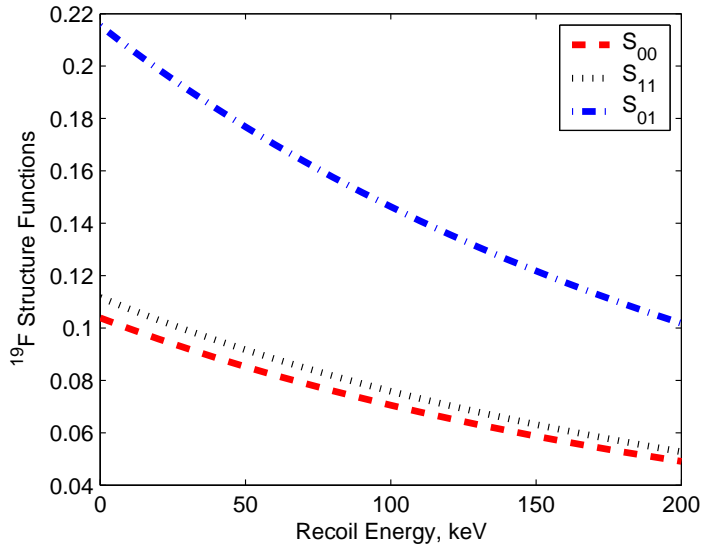}}
\end{picture}
}\end{minipage}
\hfill
\begin{minipage}[b]{0.37\textwidth}{
\caption{Structure functions
	$S^{19}_{00}$ (bottom), 
        $S^{19}_{11}$ (middle), and
	$S^{19}_{01}$ (top) for $^{18}$F 
	as a function of the recoil energy 
	$E = u b^{-2}/ M_A$
	in keV calculated by (\ref{SF-F19-pi}).
        With $v_{\max} \approx 600(700)$~km/s for the $^{18}$F target
	one has $E_{\max} \approx 140(190)$~keV.
\protect\\[10pt]}
\label{F19-bva}
}\end{minipage}
\end{figure}
      When $m_\pi\to \infty$ formulas, 
(\ref{SF-F19-pi}) pass into
(\ref{SF-F19}).
      The fluorine structure functions calculated in accordance with  
(\ref{SF-F19-pi}) are given in 
Fig.~\ref{F19-bva} as functions of the recoil energy.
      Further details can be found in the original paper
\cite{Divari:2000dc}.

\subsection{Sodium, $^{23}$Na}
         In the modern most promising scintillator dark matter detectors
(like, for example, the DAMA one) 
	 iodine is always used together with sodium in the form 
	 of large sodium iodide (NaI) crystals
(see for example, 
\cite{Bernabei:2003ky,Bernabei:2003za,%
Ahmed:2003su,Cebrian:2002vd,Yoshida:2000df}).  
	 The nucleus $^{23}$Na ($J=3/2$)
	 lies in the middle of the $sd$ shell and therefore 
	 the methods applied to other $sd$-shell nuclei
\cite{Ressell:1997kx} can be used.
         Ressell and Dean 
\cite{Ressell:1997kx} have performed calculation for $^{23}$Na 
	 exactly analogous to those done for $^{29}$Si in
\cite{Ressell:1993qm} and for $^{27}$Al in 
\cite{Engel:1995gw}, including the use of harmonic oscillator wave functions.
          The following zero-transfer-limit spin structure parameters 
	  are obtained:
$\langle {\bf S}^{23}_p\rangle = 0.2477$ and 
$\langle {\bf S}^{23}_n\rangle = 0.0198$
\cite{Engel:1995gw,Ressell:1993qm}. 
	The fits to the structure functions 
	$S^{23}_{ij}(q)$ as third order polynomials
	in $y=(qb/2)^2$ are given as follows
\cite{Ressell:1997kx}:
\begin{eqnarray}
\nonumber
S^{23}_{00}(y) &=& 0.0380 - 0.1743\, y + 0.3783\, y^2 - 0.3430\, y^3, \\
\label{Nuclear.spin.Na-23}
S^{23}_{01}(y) &=& 0.0647 - 0.3503\, y + 0.9100\, y^2 - 0.9858\, y^3, \\
\nonumber
S^{23}_{11}(y) &=& 0.0275 - 0.1696\, y + 0.5077\, y^2 - 0.6180\, y^3.
\end{eqnarray}
        Here $y$ is obviously well below $1$, 
        the functions are rather accurate to 
	$y < y_{\max} = 0.1875$ ($q_{\max}\approx 100$~MeV), 
	which corresponds to the 
	maximum halo velocity of $v_{\max} = 700$~km/s 
	and the heavy enough WIMP mass.
        Here the oscillator parameter 
	$b = 1.6864\,$fm~$ = (1/117.01)$~MeV$^{-1}$ is used.

\begin{figure}[!h] 
\begin{picture}(100,70)
\put(-68,118)
{\includegraphics{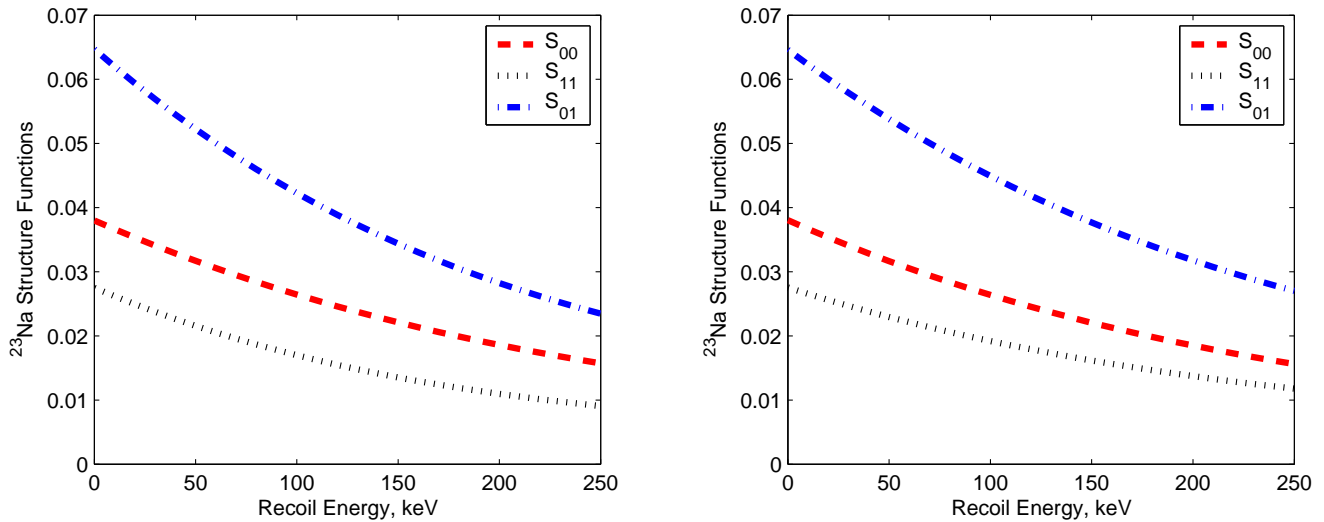}}
\end{picture}
\caption{Spin structure functions for $^{23}$Na 
	$S^{23}_{00}$ (middle),
        $S^{23}_{11}$ (bottom), and
	$S^{23}_{01}$   (top) versus the recoil energy.
	Left: $S^{23}_{ij}$ approximations from Ressell
	and Dean 
\cite{Ressell:1997kx} in accordance with  
(\ref{Nuclear.spin.Na-23}). 
	Right: $S^{23}_{ij}$ from Vergados et al.
\cite{Divari:2000dc} following 
(\ref{SF-Na23}).
        With $v_{\max} \approx 700$~km/s for $^{23}$Na target
	one has $E_{\max} \approx 230$~keV.
}
\label{Na23-bva}
\end{figure}

        Using their shell-model approach Vergados with co-authors 
        also calculated the spin structure function
	$S^{23}(q)$ for $^{23}$Na 
\cite{Divari:2000dc}.
        Their three terms of the $S^{23}(q)$ ($J=3/2$) are
\begin{eqnarray} \label{SF-Na23}
S^{23}_{00}(q) 
&=& \frac{2J+1}{16\pi} \times (0.478)\times e^{-u} 
	      \left\{  P^2_{(0,1)}(u) + P^2_{(2,1)}(u) 
	              +P^2_{(2,3)}(u) + P^2_{(4,3)}(u) 
	      \right\} , 
\\ \nonumber 
S^{23}_{11}(q) 
&=& \frac{2J+1}{16\pi} \times (0.346) \times 
              e^{-u}\left\{ Q^2_{(0,1)}(u) + Q^2_{(2,1)}(u) 
 	                    +Q^2_{(2,3)}(u) + Q^2_{(4,3)}(u) 
	              \right\} , 
\\ \nonumber
S^{23}_{01}(q) 
&=& \frac{2J+1}{8\pi}\times (0.406)\times e^{-u} 
            \left\{
             P_{(0,1)}(u) Q_{(0,1)}(u)
           + P_{(2,1)}(u) Q_{(2,1)}(u) +
\right. \\ \nonumber
&& \qquad \qquad \qquad \qquad \qquad \left.
           + P_{(2,3)}(u) Q_{(2,3)}(u)
           + P_{(4,3)}(u) Q_{(4,3)}(u)
              \right\},
\end{eqnarray} 
$${\rm       where}\quad
P_{(0,1)}(u) = 0.0477 u^2 - 0.6667 u + 1, \qquad
Q_{(0,1)}(u) = 0.0465 u^2 - 0.6667 u + 1,
$$
$$
P_{(2,1)}(u) = -0.0177\, u^2 + 0.1048\, u, \qquad
Q_{(2,1)}(u) = -0.0349\, u^2 + 0.1494\, u,
$$
$$
P_{(2,3)}(u) = -0.0767\, u^2 + 0.6092\, u, \qquad
Q_{(2,3)}(u) = -0.0894\, u^2 + 0.7405\, u,
$$
$$
P_{(4,3)}(u) = 0.0221\, u^2, \qquad
Q_{(4,3)}(u) = 0.0287\, u^2 .
$$
       The dimensionless variable $u=(qb)^2/2\equiv 2y$, 
       $b=1.00A^{1/6} = 1.69\,$fm 
       is the oscillator size parameter for $^{23}$Na. 
       Relevant spin static characteristics of 
\cite{Divari:2000dc}\
$\langle {\bf S}^{23}_p\rangle = 0.2477$ and 
$\langle {\bf S}^{23}_n\rangle = 0.0199$
        are almost exactly equal to ones from 
\cite{Ressell:1997kx}.
        In 
Fig.~\ref{Na23-bva} the recoil energy dependence of the 
        spin structure functions calculated for $^{23}$Na by 
(\ref{Nuclear.spin.Na-23}) and 
(\ref{SF-Na23}) is presented. 
       One can see from 
Fig.~\ref{Na23-bva} that the isoscalar spin functions $S^{23}_{00}$ have
        practically the same behavior in the left and right panels, 
	but, despite the same normalization at $q=0$,
	the isovector $S^{23}_{11}$ and mixed $S^{23}_{01}$ 
	spin functions 
	from Vergados et al.
\cite{Divari:2000dc} (in the right panel) 
        are systematically a bit larger than the 
	$S^{23}_{11}$ and $S^{23}_{01}$ 
	functions from Ressell and Dean
\cite{Ressell:1997kx} (in the left panel of 
Fig.~\ref{Na23-bva}).

\subsection{Aluminum, $^{27}$Al}
	The $^{27}$Al nucleus is one of the active ingredients 
	in a very high-resolution 
	and low-threshold 
	sapphire-crystal (Al$_2$O$_3$) CRESST detector
\cite{Angloher:2003cg}. 
        Engel, Ressell, Towner, and Ormand in 
\cite{Engel:1995gw} carried out
	calculations with the Lanczos m-scheme shell-model code CRUNCHER
\cite{Resler:1988aa}. 
	The m-scheme basis for $^{27}$Al contains 80115 Slater determinants.  
	The agreement of the calculated spectrum with that
	measured in $^{27}$Al is very good.
	A similar calculation for $^{29}$Si was fulfilled by Ressell et al. 
\cite{Ressell:1993qm}.
	In the shell model, the expectation value of any 
	one-body operator and therefore the structure functions
	can be easily calculated.
	To evaluate the $S^{27}(q)$
        Engel, Ressell, Towner, and Ormand used 
	a nuclear mean-field potential
	which is closer to the Woods-Saxon rather 
	than the harmonic oscillator form.
	The length parameter $b=1.73$~fm for the oscillator functions and
	the standard parameters for the Woods-Saxon potential were used
\cite{Engel:1995gw} in  calculation of 
	the structure functions. 
\begin{figure}[h!] 
\begin{minipage}[b]{0.45\textwidth}
{\begin{picture}(45,75)
\put(-55,-87){\includegraphics{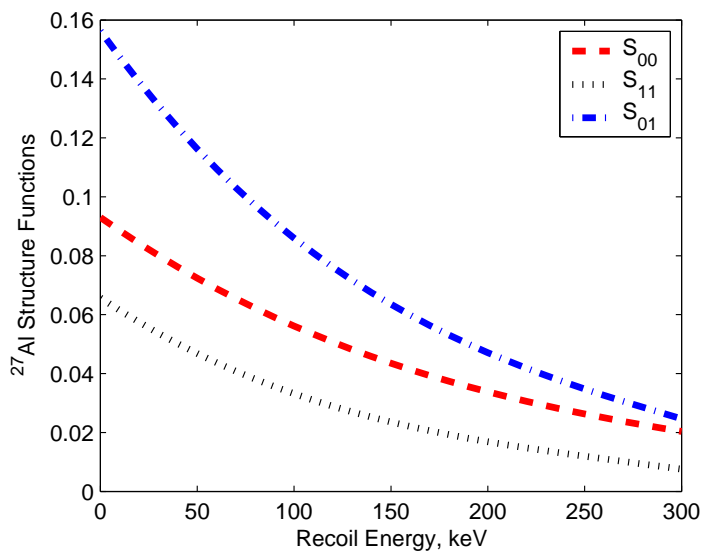}}
\end{picture}
}\end{minipage}
\hfill
\begin{minipage}[b]{0.37\textwidth}{
\caption{Structure functions
	$S^{27}_{00}$ (middle), $S^{27}_{11}$ (bottom), and
	$S^{27}_{01}$ (top) for $^{27}$Al 
	as a function of the recoil energy calculated by
(\ref{Nuclear.spin.AL-27}).
	For $^{27}$Al and the WIMP maximal velocity of 700 km/s
	the maximum  momentum
	transfer is $q_{\max} \approx  117$~MeV and $E_{\max} \approx 270$~keV.
}
\label{mica-Al27}
}\end{minipage}
\end{figure}
        The most accurate $^{27}$Al structure functions of 
\cite{Engel:1995gw}
	with the Woods-Saxon single-particle wave functions
        can be given to a high accuracy by the third-order polynomials
\begin{eqnarray}\nonumber
S^{27}_{00}(q)&=& 0.0930 - 0.4721\, y + 1.0600\, y^2 -1.0115\, y^3,   \\
\label{Nuclear.spin.AL-27}
S^{27}_{11}(q)&=& 0.0657 - 0.4498\, y + 1.3504\, y^2 -1.6851\, y^3,  \\
S^{27}_{01}(q)&=& 0.1563 - 0.9360\, y + 2.4578\, y^2 -2.7262\, y^3,
\nonumber
\end{eqnarray}
	where $y=(bq/2)^2$.
Figure~\ref{mica-Al27} 
        shows these functions $S^{27}_{ij}$ for aluminum 
	versus the recoil energy (see formulas 
(\ref{Nuclear.spin.AL-27})). 
	These three functions allow one to calculate 
	the spin-dependent cross sections for the neutralino 
	interaction with the $^{27}$Al target
\cite{Engel:1995gw}.

\subsection{Silicon, $^{29}$Si}
         The first accurate calculation 
	 of the $q$-dependence of the spin structure functions 
	 for WIMP scattering off  $^{29}$Si 
	 was carried out by Ressell et al.
\cite{Ressell:1993qm}.
	 With a reasonable two-body interaction Hamiltonian 
	 Ressell et al. calculated nuclear wave functions 
	 in an appropriate model space.
	 The accuracy of those wave functions 
	 was checked by comparison of the calculated 
	 excited state energy spectrum,  magnetic moments, 
	 and spectroscopic factors with the experimental observables.
	 The checked ground state wave functions 
	 were used for further calculations of
	 static and more complicated 
	 finite momentum nuclear matrix elements
	 involved in DM search calculations.
         Finite momentum transfer matrix elements and
        cross sections for the spin-dependent elastic scattering  
	of neutralinos from $^{29}$Si and $^{73}$Ge
\cite{Ressell:1993qm} were evaluated in that shell-model scheme.  
	 Both these isotopes have a great number of 
	 configuration mixing and require 
	 very large model spaces.

	 In particular, for evaluation of the wave functions for 
	 $^{29}$Si the universal $sd$ shell-model interaction of 
	 Wildenthal (see, for example, 
\cite{Brown:1985aa,Brown:1988aa}) 
	 in a full $sd$  shell-model space was used
\cite{Ressell:1993qm} 
         and calculations were carried out
	 with the Lanczos m-scheme shell-model code CRUNCHER
\cite{Resler:1988aa}. 
	 As in the case of $^{27}$Al,
\cite{Engel:1995gw} 
	 the m-scheme basis for $^{29}$Si 
	 contains 80115 Slater determinants.  
	 For static spin matrix elements  the following 
	 values were obtained:
	$\langle{\bf S}^{29}_p\rangle = -0.002$ and 
	$\langle{\bf S}^{29}_n\rangle = 0.13$ (given in Table 4 of 
\cite{Bednyakov:2004xq}). 

\begin{figure}[!h] 
\begin{picture}(100,65)
\put(-67,118)
{\includegraphics{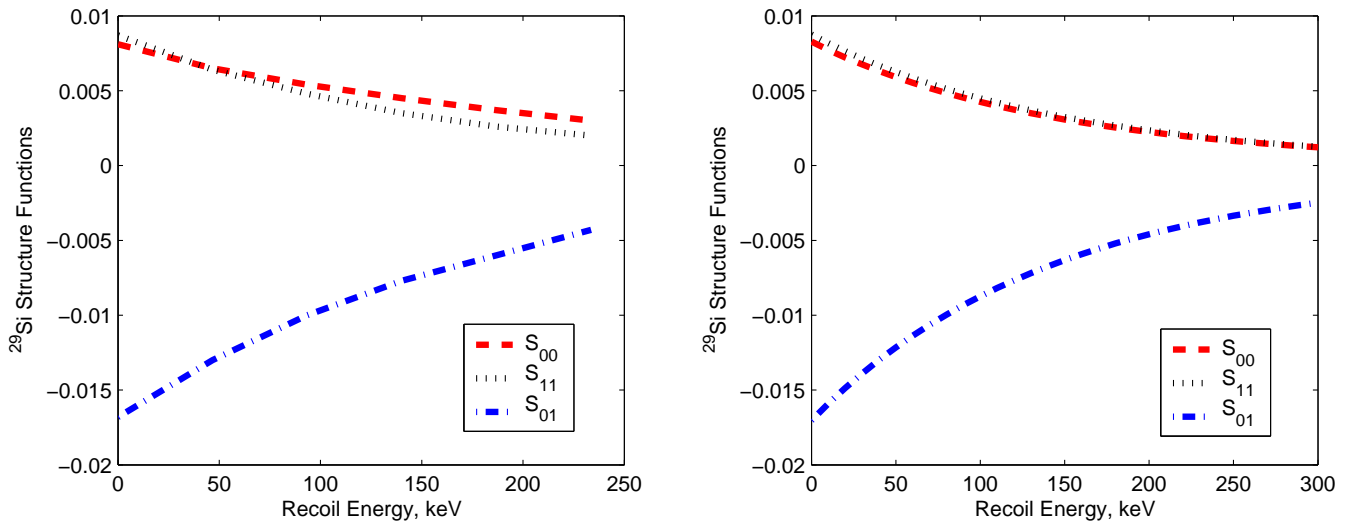}}
\end{picture}
\caption{Spin structure functions 
         $S^{29}_{00}$ (top)  
         $S^{29}_{11}$ (middle), and
         $S^{29}_{01}$ (bottom) for $^{29}$Si 
        as a function of the recoil energy.
	Left: results of Ressell et al. 
\protect\cite{Ressell:1993qm}. 
	Right: these structure functions from Vergados et al.
\cite{Divari:2000dc} 
        following equations
(\ref{SF-Si29-pi}).
        With $v_{\max} \approx 700$~km/s, for $^{29}$Si target
	one has $E_{\max} \approx 290$~keV.
}
\label{S29-bva}
\end{figure}

Figure~\ref{S29-bva} (left panel)
        shows the recoil energy dependence of the spin 
	partial structure functions 
	$S^{29}_{00}$, $S^{29}_{11}$ and $S^{29}_{01}$ 
	for $^{29}$Si originally calculated in  
 \cite{Ressell:1993qm} as a function of 
        $y=(bq/2)^2$ with 
	the oscillator parameter $b=1.75~$fm for $^{29}$Si. 
	With $v_{\max}= 600 (700)$~km/s the maximal momentum transfer is 
	$q_{\max}= 0.108 (126)~$GeV, $y_{\max}= 0.23 (0.31)$ and
	$E_{\max}= 213 (290)$~keV.

	For user's convenience, in their earlier work 
 \cite{Ressell:1993qm} Ressell et al. 
	obtained rather simple parameterizations of the {\em full}\/ 
	spin structure function $S^{29}(q)$ (quenching is included)
\begin{equation}
\label{Ressell-fit-Si}
S_{\rm fit}^{29}(y) 
= 0.00818\, \big( a_0^2 e^{-4.428 y} + 1.06 a_1^2 e^{-6.264 y} 
                - 2.06 a_0 a_1 e^{-5.413 y} \big),
\end{equation}
        which provides a highly accurate fit to $S^{29}(y)$ for $y < 0.15$.  
	For $y\ge 0.15$ this parameterization begins seriously 
	underestimate the true value of $S^{29}(y)$
 \cite{Ressell:1993qm}.
       The lack of validity of 
(\ref{Ressell-fit-Si}) 
       at high $y$ (or $q$) is not a serious problem for WIMP masses up to 
       about 600 GeV/$c^2$.
       Anyway, when $a_0\approx a_1$, parameterization 
(\ref{Ressell-fit-Si}) does not reproduce accurately the full result of 
\cite{Ressell:1993qm} for all $y$ given in the left panel of  
Fig.~\ref{S29-bva}. 

 	Vergados with co-authors also calculated the spin structure function
	$S^{29}(q)$ for $^{29}$Si in their approach of 
\cite{Divari:2000dc}.
        The pure isoscalar, $S^{29}_{00}$, pure isovector, $S^{29}_{11}$, 
	and interference, $S^{29}_{01}$ terms of the 
	silicon structure function
(\ref{Definitions.spin.decomposition}) 
	$S^{29}(q)$ ($J^\pi=1/2^+$) are given in the form
\begin{eqnarray}
\nonumber
S^{29}_{00}(q) 
&=& \frac{2J+1}{16\pi} \times (0.208)\times 
              e^{-u}\left\{P^2_{(0,1)}(u) + P^2_{(2,1)}(u) \right\} , 
\\ \label{SF-Si29}
S^{29}_{11}(q) 
&=& \frac{2J+1}{16\pi} \times (0.220) \times 
              e^{-u}\left\{Q^2_{(0,1)}(u) + Q^2_{(2,1)}(u) \right\} , 
\\ 
\nonumber
S^{29}_{01}(q) 
&=& \frac{2J+1}{8\pi}\times (-0.214)\times 
            e^{-u}\left\{
             P_{(0,1)}(u) Q_{(0,1)}(u)
           + P_{(2,1)}(u) Q_{(2,1)}(u)
              \right\}.
\end{eqnarray} 
$${\rm       where}\quad
P_{(0,1)}(u) = 0.2843 
                     \, u^2 - 0.6667\, u + 1, \qquad
Q_{(0,1)}(u) = 0.2710 
                     \, u^2 - 0.6667\, u + 1,
$$
$$
P_{(2,1)}(u) = -0.0567\, u^2 + 0.4566\, u, \qquad
Q_{(2,1)}(u) = -0.0621\, u^2 + 0.4680\, u.
$$
       Here $u=(qb)^2/2 \equiv 2y$. 
       For a rather large momentum transfer, 
       in full analogy with $^{19}$F (see, formulas
(\ref{SF-F19-pi})), one has 
\cite{Divari:2000dc}:
\begin{eqnarray}
\nonumber
S^{29}_{00}(q) &=& 
\frac{0.208}{8\pi}\,e^{-u}\left\{ P^2_{(0,1)}(u) (1+\beta_0(u)) 
                    +  P^2_{(2,1)}(u) (1+\beta_2(u)) 
            - 2\beta_{02}(u) P_{(0,1)}(u) P_{(2,1)}(u)\right\} , 
\\ \nonumber
S^{29}_{11}(q) &=& 
  \frac{0.220}{8\pi} \,
              e^{-u}\left\{ Q^2_{(0,1)}(u) (1+\beta_0(u)) 
                          + Q^2_{(2,1)}(u) (1+\beta_2(u)) 
	- 2 \beta_{02}(u) \, Q_{(0,1)}(u) Q_{(2,1)}(u)\right\} , 
\\ \nonumber
S^{29}_{01}(q) &=& 
       \frac{-0.214}{4\pi}\,e^{-u}\left\{
             P_{(0,1)}(u) Q_{(0,1)}(u) (1+\beta_0(u)) 
           + P_{(2,1)}(u) Q_{(2,1)}(u) (1+\beta_2(u)) \right. \\
\label{SF-Si29-pi}
   && \qquad\qquad \qquad\qquad 
      \left. -\beta_{02}(u)\, (
               P_{(0,1)}(u) Q_{(2,1)}(u)+Q_{(0,1)}(u) P_{(2,1)}(u) )
\right\}, 
\end{eqnarray}
     where 
     $\beta_0(u)$, $\beta_2(u)$, and $\beta_{02}(u)$ are defined in 
(\ref{Pion-finite-mass}). 
Figure~\ref{S29-bva} (right panel) presents structure functions 
        for $^{29}$Si from Vergados et al.~%
\cite{Divari:2000dc} following formulas
(\ref{SF-Si29-pi}).
	Both sets of spin structure functions for $^{29}$Si   
	given in the left 
\cite{Ressell:1993qm} and right 
\cite{Divari:2000dc} panels of 
Fig.~\ref{S29-bva} are very similar.

\subsection{Potassium, $^{39}$K}
	In the case of $^{39}$K the shell-model diagonalization needed for
        the calculation of the nuclear spin matrix
        elements requires severe truncations to the active model space. 
        The problem is that $^{39}$K is near the boundary between the 
	{\it sd} and {\it pf} shells and
	excitations of particles into higher shells can have significant
	effects that are often not well simulated by effective operators.
        Thus, for this nucleus Engel, Ressell, Towner and Ormand 
\cite{Engel:1995gw} 
        used  an alternative scheme based on perturbation theory 
	for the evaluation of spin matrix elements. 	
        The authors considered two different residual interactions. 
	One 
	is related to the one-boson-exchange potential of the Bonn type,
	but it is limited only to four or five important meson exchanges.  
	The resulting interaction has a weak tensor-force component 
	typical of Bonn potentials.  
	The other 
	is represented by full G-matrix elements of the Paris potential 
	parameterized in terms of sums over Yukawa functions 
	of various ranges and strengths. 
	This interaction 
	exhibits a strong tensor force.
        The quality of the wave functions obtained was judged in terms  
	of magnetic moments and Gamow-Teller matrix elements, 
	including meson-exchange currents, isobar
	currents, and other relativistic effects. 
        The magnetic moments calculated with the help of both interactions 
        differ only slightly from each other and showed good agreement
        with the corresponding  experimental values.  
        The same nuclear wave functions of $^{39}$K were also used
	for the calculation of 
	$\langle{\bf S}^{39}_p\rangle$ and 
	$\langle{\bf S}^{39}_n\rangle$ 
	and the structure function $S^{39}(q)$
\cite{Engel:1995gw}. 
        Three different  sets of spin structure functions for $^{39}$K
	were considered in 
\cite{Engel:1995gw}.
         They are the ``single-particle'' functions, 
	 the full functions obtained from the 
	 modified Bonn interaction, and the full
	 functions obtained from the Paris-based G-matrix.

\begin{figure}[h!] 
\begin{minipage}[b]{0.47\textwidth}
{\begin{picture}(50,70)
\put(-50,-83){\includegraphics{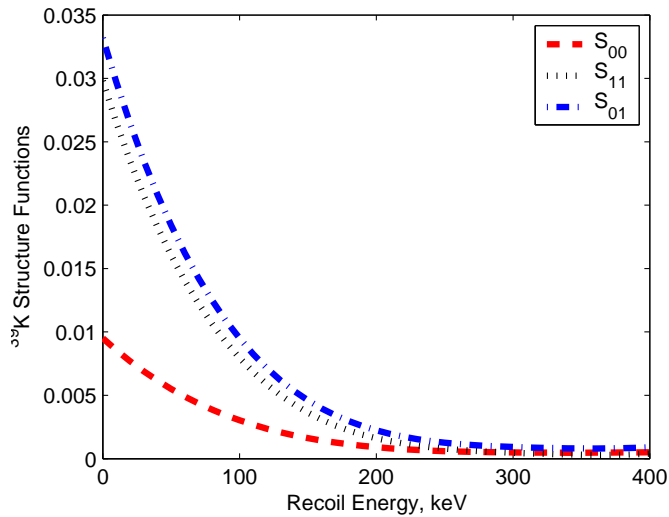}}
\end{picture}
}\end{minipage}
\hfill
\begin{minipage}[b]{0.40\textwidth}{
\caption{$^{39}$K spin structure functions
         $S^{39}_{00}$ (top), 
         $S^{39}_{11}$ (middle), and
         $S^{39}_{01}$ (bottom) versus 
         the recoil energy, calculated by 
(\ref{SF-39K}) from Engel, Ressell, Towner and Ormand 
\protect\cite{Engel:1995gw}.
         With $v_{\max}=700$~km/s, for $^{39}$K\/  one has 
	 $q_{\max} = 169$~MeV, $E_{\max} = 390$~keV.
\protect\\
}
\label{S39-bva} 
}\end{minipage}
\end{figure} 

	There is rather strong reduction of 
	$S^{39}_{ij}$ in comparison with their single-particle values.  
	The strongest reduction is in $S^{39}_{00}(q)$, 
	which is reduced to 25\% and 20\% of the
	single-particle value for the two residual interactions.  
	The preferred choice (corresponding to 
	the Paris-based G-matrix approach) 
	can be rather accurately reproduced by the
	following fourth-order polynomials in $y=(bq/2)^2$
	(with $b=1.84$~fm for $^{39}$K):
\begin{eqnarray}\nonumber
S^{39}_{00}(q)&=& 0.0095 - 0.0620\,y + 0.1628\,y^2 - 0.1943\,y^3 + 0.0891\,y^4,
\\ \label{SF-39K}
S^{39}_{11}(q)&=& 0.0298 - 0.2176\,y + 0.6236\,y^2 - 0.8144\,y^3 + 0.4050\,y^4, 
\\ \nonumber
S^{39}_{01}(q)&=& 0.0332 - 0.2319\,y + 0.6385\,y^2 - 0.7985\,y^3 + 0.3810\,y^4.
\end{eqnarray}
	The spin structure functions 
(\ref{SF-39K}) for $^{39}$K are given in 
Fig.~\ref{S39-bva}.
           These structure functions allow one to
	   interpret experiments that can look for 
	   tracks due to the interaction of
	   dark-matter particles with nuclei in ancient mica
\cite{Engel:1995gw}.

\subsection{Germanium, $^{73}$Ge}
	Germanium isotopes, 
	especially large-spin ($J=9/2$) ${}^{73}$Ge, are considered to be 
	the most promising material for a real long-running experiment 
	aimed at direct dark matter search. 
	The first experiment with a pure (enriched) ${}^{73}$Ge target 
	was successfully performed at Gran Sasso by the HDMS collaboration
\cite{Klapdor-Kleingrothaus:2002pg,Klapdor-Kleingrothaus:2005rn}.
	However, there are fundamental difficulties in describing 
	the spin content 
	of ${}^{73}$Ge due to its complicated collective structure.

         The first accurate calculation 
	 of the $q$-dependence of the spin structure functions 
	 for WIMP scattering off  $^{73}$Ge 
	 was carried out by Ressell at al. in 
\cite{Ressell:1993qm} (together with $^{29}$Si).
	An equally comprehensive calculation 
	was realized by Dimitrov, Engel and Pittel 
\cite{Dimitrov:1995gc}. 
        These authors obtained
        significantly different (and improved) results in comparison 
        with other studies and they argue 
	that their results are more reliable than the previous ones.

	For the study of $^{73}$Ge Ressell et al.
\cite{Ressell:1993qm} chose the 
	Petrovich-McManus-Madsen-Atkinson interaction
\cite{Petrovich:1969aa}, 
	which is a reasonable approximation to a full G-matrix calculation. 
	This interaction proved to be both adequate and tractable 
	in shell model applications. 
	Two different model spaces were considered.
	The  ``small'' space was determined by an m-scheme basis 
	dimension of 24731 Slater determinants. 
	The ``large'' space allowed much more excitations 
	with an m-scheme basis dimension of 117137 Slater determinants. 
	Despite fairly large size of the bases, 
	rather severe truncations in the space were implemented. 
	The small space is the smallest one in which it is possible
	to obtain agreement with the experimental spectrum energy levels. 
	The dimension of the large basis  was limited by the  computer 
	time and the memory storing constraints
\cite{Ressell:1993qm}. 
	No phenomenological interaction has been developed for 
	Ge-like nuclei and fairly 
	severe truncations to the model space have to be 
	imposed to obtain manageable dimensions.
        The large model space wave function of $^{73}$Ge led
        to an improved description of the ground state expectation
        values, in particular of the value of the magnetic moment,
        in comparison with previous estimates. 
	The calculated magnetic moment $\mu$ from 
\cite{Ressell:1993qm} exceeds the experimental value, but the authors stressed 
	that the same quenching of both $\mu$ 
	and the Gammov-Teller spin matrix elements was almost 
	universally required in shell model calculations of 
        all heavy nuclei. 
	Assuming the isovector spin quenching factor  
	to be $0.833$, agreement with the measured $\mu$ is obtained.
	In principle, it is not obvious that quenching is really
	needed in neutralino-$^{73}$Ge scattering but if so, 
	Ressell et al. believed that the correct answer might
	be in the range between the quenched and unquenched values. 
	It was found (see Table~6 of 
\cite{Bednyakov:2004xq})
	that the zero-momentum-transfer spin-neutron 
	matrix element 	$\langle{\bf S}^{73}_{n}\rangle$ of
        $^{73}$Ge was a factor of 2 larger than the previous predictions.
	Thus,
	even if quenching is assumed, the calculated scattering rate 
	is about twice as large as any of the estimates obtained before.
        Therefore Ressell et al. predicted 
	a higher sensitivity for germanium dark matter detectors
\cite{Ressell:1993qm}.

\begin{figure}[!h] 
\begin{picture}(100,70)
\put(-67,120)
{\includegraphics{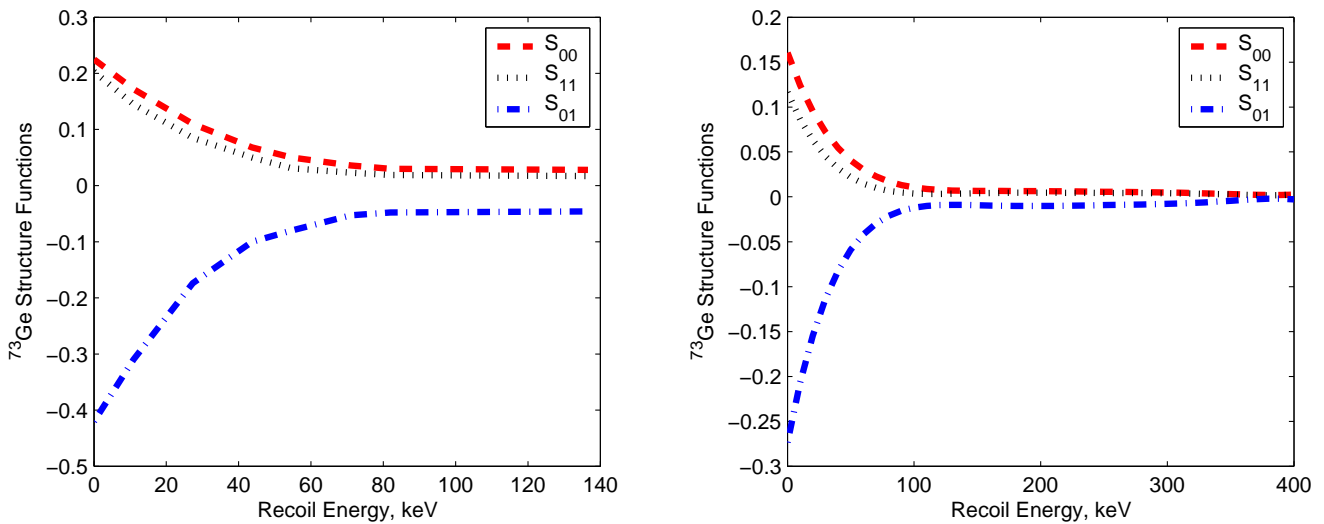}}
\end{picture}
\caption{Spin structure functions 
         $S^{73}_{00}$ (top)  
         $S^{73}_{11}$ (middle), and
         $S^{73}_{01}$ (bottom) for $^{73}$Ge 
        as a function of the recoil energy.
	Left: results of Ressell et al. 
\cite{Ressell:1993qm}.
	Right: the same
	structure functions, but from the ``hybrid'' method of
        Dimitrov, Engel and Pittel
\cite{Dimitrov:1995gc} 
        following equations
(\ref{Dimitrov-Ge73-SF}).
        With $v_{\max} \approx 600$~km/s, for the $^{73}$Ge target
	one has $E_{\max} \approx 537$~keV.
}
\label{S73-bva}
\end{figure}

Figure~\ref{S73-bva} (left panel) 
        shows the recoil energy dependence of the partial structure functions 
	$S^{73}_{00}$, $S^{73}_{11}$ and $S^{73}_{01}$ 
	for $^{73}$Ge calculated by Ressell et al. in 
 \cite{Ressell:1993qm} in terms of $y=(bq/2)^2$. 
        For $^{73}$Ge the oscillator parameter $b=2.04~$fm, 
	the maximal momentum transfer ($v_{\max}=600$~km/s) is 
	$q_{\max}=0.271~$GeV 
	and $E_{\max} = 537$~keV.
	As in the case of $^{29}$Si (see expression 
(\ref{Ressell-fit-Si})) Ressell et al. 
	gave rather simple parameterizations of the
	{\em full}\/ spin structure function $S^{73}(q)$
	(where quenching is included and which is valid for $y < 0.2$):
\begin{equation}
\label{Ressell-fit-Ge}
S_{\rm fit}^{73}(y) = 0.20313\, \big( 1.102 a_0^2 e^{-7.468 y} + a_1^2
e^{-8.856 y} - 2.099 a_0 a_1 e^{-8.191 y} \big).
\end{equation}
       The lack of validity of 
(\ref{Ressell-fit-Ge}) for $^{73}$Ge
       at high $y$ (or $q$) is not a problem for WIMP masses up to 
       about 600 GeV$/c^2$.
       When $a_0\approx a_1$,  parameterization 
(\ref{Ressell-fit-Ge}) does not reproduce accurately 
       the full result for all $y$ 
 \cite{Ressell:1993qm}.

	A different sophisticated approach to evaluation of the spin 
	structure of $^{73}$Ge was considered by Dimitrov, Engel and Pittel in
\cite{Dimitrov:1995gc}.
	It relies on the idea of mixing variationally determined
	Slater determinants, in which symmetries are broken 
	but restored either before or after variation.  
        This approach is described in detail in
\cite{Dimitrov:1994aa}.  
	In the calculation of
\cite{Dimitrov:1995gc} 
	the symmetries broken in the intrinsic states are those 
	associated with rotational invariance, parity, and axial shape.  
	The hybrid procedure used restores axial symmetry,
	parity invariance, and approximate rotational invariance prior to
	the variation of each intrinsic state.   
	Subsequently, before mixing the intrinsic states the
	rotational invariance is fully restored.
	The procedure allows fully triaxial Slater
	determinants at the expense of particle-number breaking.  
	The results of 
\cite{Dimitrov:1994aa} indicate that the trading of number 
	nonconservation for triaxiality is a good idea, 
	despite the apparent loss of pairing correlations
	traditionally associated with the former.  
	Pairing forces evidently induce effective triaxiality.  
	The numerical results
\cite{Dimitrov:1994aa} show that the approach is
	accurate and efficient for describing even-even systems 
	while also providing reliable reproduction 
	of the collective dynamics of odd-mass systems
\cite{Dimitrov:1995gc}.

	For ${}^{73}$Ge the calculations in
\cite{Dimitrov:1995gc} were performed  
        by assuming, both for protons and neutrons, a single-particle model 
	consisting of the full $0f,1p$ shell and  
	the $0g_{9/2}$ and $0g_{7/2}$ levels.
	The main idea was to include all of the single-particle orbits that
	could play an important role in reproducing 
	low-energy properties of ${}^{73}$Ge 
\cite{Dimitrov:1995gc}.
	It is well-known that a crucial 
        ingredient in any realistic nuclear-structure 
	calculation is the appropriate form of the nuclear Hamiltonian.  
	The one- and two-body parts of the Hamiltonian 
	have to be compatible with each other as well as with the model space.
	This is difficult to achieve because 
	microscopic two-body interactions, derived for example from a
	G-matrix, include monopole pieces that are unable 
	to describe the movement of spherical single-particle 
	levels as one passes from the beginning to the end of a shell.
	A  procedure proposed for avoiding this problem 
	consists basically in removing all monopole components 
	from the two-body interaction and 
	shifting their effects to the single-particle energies.  
	This procedure was used by Dimitrov, Engel and Pittel 
\cite{Dimitrov:1995gc} --- their two-body
	force was a fit to the Paris-potential G-matrix
	modified as described above.
	The calculated ground-state magnetic dipole moment is in
	good agreement with the experimental value.  
	Ressell et al. 
\cite{Ressell:1993qm} 
	in their large-space shell-model
	calculation were able to reduce $\mu$ significantly to 
	$-1.239 \mu_N$ 
	(with experimental value $-0.879 \mu_N$)
	but could not account for the remaining difference.  
	On the contrary, the calculation of Dimitrov, Engel and Pittel, 
	despite the small number of intrinsic states, 
	contains the full quenching required by the experimental data
\cite{Dimitrov:1995gc}.
	By making a comparison with the results of Ressell et al. 
\cite{Ressell:1993qm}, 
        significant disagreement is found for the neutron spin again. 
	The calculated value  of Dimitrov, Engel and Pittel 
	is significantly smaller ($-0.920\mu_N$).  
	The differences in the spin contributions, unlike those in the 
	orbital angular momenta, strongly affect the WIMP-Ge 
	scattering cross sections. 
	Thus, contrary to Ressell et al. 
\cite{Ressell:1993qm}, 
	no significant increase can be
	expected in the neutralino-${}^{73}$Ge scattering rate
	in accordance with
\cite{Dimitrov:1995gc}. 

	The advantage of Dimitrov, Engel and Pittel's approach 
	to calculation of neutralino cross sections 
	is that it correctly represents the spin structure, 
	requires neither quenching at $q=0$ nor arbitrary assumptions about 
	the form factor behavior at $q \neq 0$
\cite{Dimitrov:1995gc}.  
Figure~\ref{S73-bva} (right panel) 
        shows partial structure functions of Dimitrov, Engel and Pittel
\cite{Dimitrov:1995gc}
        that determine the spin-dependent cross sections
	of elastic neutralino scattering off ${}^{73}$Ge. 

      Comparing the results for $S^{73}_{00}(q)$ (the pure
      isoscalar form factor) and $S^{73}_{11}(q)$ (the isovector form factor) 
      with the corresponding large-space results of Ressell et al.
\cite{Ressell:1993qm} given in  
Fig.~\ref{S73-bva} (left panel) 
       one can conclude that both $S^{73}_{00}(q)$ and $S^{73}_{11}(q)$
       of Dimitrov, Engel and Pittel
       are reduced relative to the curves of Ressell et al.
\cite{Ressell:1993qm}.

       The polynomial fits, in terms of $y=(b q/2)^2$ with 
       $b=2.04$~fm being the oscillator parameter for ${}^{73}$Ge, which
       well represent the structure functions 
in Fig.~\ref{S73-bva} (right panel), 
       have the forms of the following sixth-order polynomials
\cite{Dimitrov:1995gc}:
\begin{eqnarray}
\label{Dimitrov-Ge73-SF}
S^{73}_{00}(y) &=& 0.1606 - 1.1052 y + 3.2320y^2 - 4.9245 y^3
	     +4.1229 y^4 - 1.8016 y^5 + 0.3211 y^6, \\
\nonumber
S^{73}_{11}(y) &=& 0.1164 - 0.9228~y + 2.9753y^2 - 4.8709~y^3
	     +4.3099 y^4 - 1.9661 y^5 + 0.3624 y^6, \\
S^{73}_{01}(y) &=&-0.2736 + 2.0374y  - 6.2803~y^2 + 9.9426 y^3
	     -8.5710 y^4 + 3.8310 y^5 - 0.6948~y^6.
\nonumber
\end{eqnarray}
       Here we have to stress that expressions 
(\ref{Dimitrov-Ge73-SF}) are valid only for $y<1$ (the smaller $y$ the 
       better the accuracy).  
       At the same time for ${}^{73}$Ge, when 
       the maximal WIMP velocity $v_{\max}=600$~km/s, 
       one already has the maximal possible momentum transfer 
       $q_{\max}=0.271~$GeV ($E_{\max} = 537$~keV, $q_{\max}b=2.08$) 
       and $y_{\max}=1.96>1$. 
       Therefore formulas
(\ref{Dimitrov-Ge73-SF}) are obviously {\em not}\/ valid 
       over the full range of relevant momenta transfer.
       From 
Fig.~\ref{S73-bva} (right panel) it seems, perhaps, 
       rather safe in practice to put all $S^{73}_{ij}(y)=0$ for $y\ge 1$.

	Finally, it will be a very hard task to substantially improve
	the calculations of Ressell et al. 
\cite{Ressell:1993qm} and Dimitrov, Engel and Pittel
\cite{Dimitrov:1995gc} for these spin matrix elements.
        Nevertheless, Dimitrov, Engel and Pittel have proposed 
\cite{Dimitrov:1995gc}
	several possible improvements to their analysis in the future.  
	The most important is an explicit incorporation 
	of the remaining orbits from the $2s$,$1d$,$0g$ shell, 
	whose effects were treated very roughly in 
\cite{Dimitrov:1995gc}. 
        This should enable one
	to remove some of the arbitrariness in the single-particle
	energies that resulted from 
	incomplete treatment of parity-mixing effects, etc
(for details see original paper
\cite{Dimitrov:1995gc}).

\subsection{Niobium, $^{93}$Nb}
	The niobium isotope $^{93}$Nb is an odd-group proton nucleus 
	with a large ground-state angular momentum $J=9/2$. 
	It is a heavy enough nucleus, which can be represented by 
	a basic shell-model space corresponding to three 
	protons in the $1p_{1/2}$ or $0g_{9/2}$ levels and 
	two neutrons in the $1d_{5/2}$ level.
        This model space was considered by Engel, Pittel, Ormand and Vogel  
\cite{Engel:1992qb}. 
	In order to obtain better agreement with the experiment
        the authors extended it including in the ``large'' basis  
	all states in which one proton or one neutron is excited 
        from the above-mentioned ``small'' space 
	to any level in the $sdg$-shell.
	The resulting space contains about 2700 states.	
\begin{figure}[h!] 
\begin{minipage}[b]{0.57\textwidth}
{\begin{picture}(50,75)
\put(-57,-80){\includegraphics{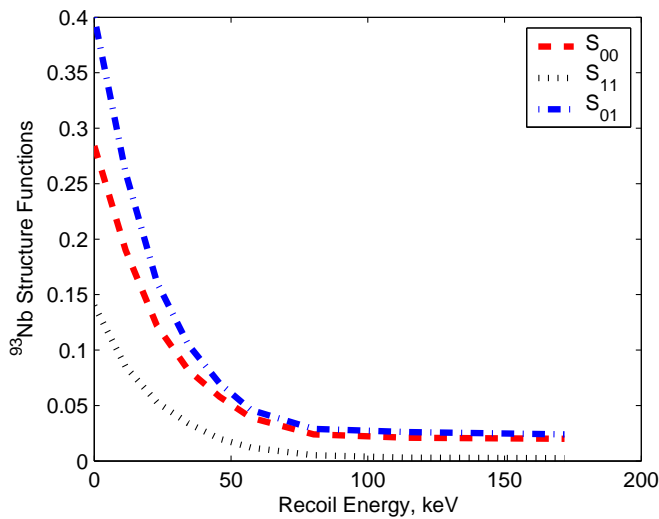}}
\end{picture}
}\end{minipage}
\hfill
\begin{minipage}[b]{0.35\textwidth}{
\caption{Partial structure functions 
        $S^{93}_{00}(q)$ (middle), 
        $S^{93}_{01}(q)$ (top) and 
        $S^{93}_{11}(q)$ (bottom) 
	in ${}^{93}$Nb as a function of the recoil energy 
	obtained from paper of Engel, Pittel, Ormand and Vogel 
\cite{Engel:1992qb}.
       Note that for ${}^{93}$Nb, when 
       the maximal WIMP velocity $v_{\max}=600$~km/s, 
       one has 
       $q_{\max}=0.345$~GeV/$c$ (and $q^2_{\max}=0.12$~GeV$^2$/$c^2$), 
       $E_{\max} = 684$~keV
       and $y_{\max}=3.48$. 
}
\label{S93-bva} 
}\end{minipage}
\end{figure} 
	They ended up with the 
	nuclear spin matrix elements
\cite{Engel:1992qb}
	$\langle{\bf S}^{93}_{p}\rangle =  0.46$ and 
	$\langle{\bf S}^{93}_{n}\rangle =  0.08$
\footnote{In Table 6 of 
\cite{Bednyakov:2004xq} these values should 
        correspond to the large shell-model spaces.}.
        With relation 
(\ref{SF-normalization}) these values can be used to check
	normalization of the 
	partial spin structure functions $S^{93}_{ij}(q)$ 
        which are available only graphically from Fig.~3 of 
\cite{Engel:1992qb} (and are given here in 
Table~\ref{S93-digitaized}). 
       In 
Fig.~\ref{S93-bva} we present recoil energy dependence of the 
       $S^{93}_{ij}$ structure functions from 
Table~\ref{S93-digitaized}.
\begin{table}[h!]
\caption{Tabulated partial spin structure functions $S^{93}_{ij}(q)$ from 
\protect\cite{Engel:1992qb} with $q^2$ in GeV$^2$/$c^2$.}
\label{S93-digitaized}
\begin{tabular}{|r|r|r|r|r|r|r|r|r|r|} \hline 
$q^2$ 
&~~0	  &~~0.002 &~~0.004 &~~0.006 &~~0.008 &~~0.01&~~0.014 &~~0.02 &~~0.03  \\
\hline \vphantom{\Huge Y}
$S^{93}_{01}(q)$ 
&~~0.4   &~~0.26  &~~0.161 & 0.105 & 0.07  & 0.046 & 0.029 & 0.026 & 0.024 \\
$S^{93}_{00}(q)$ 
&~~0.284 &~~0.19  &~~0.122 & 0.082 & 0.058 & 0.039 & 0.024 & 0.021 & 0.02  \\
$S^{93}_{11}(q)$ 
&~~0.14  &~~0.085 &~~0.053 & 0.034 & 0.02  &~~0.012 &~~0.0052&~~0.003 &~~0.0025\\
\hline
\end{tabular}
\end{table}

        An interesting observation concerning interplay between 
	spin-dependent and spin-independent $q$-dependencies
	of nuclear structure functions is given by Engel in
\cite{Engel:1992qb}. 
        He noted that
        only nucleons near the Fermi surface
          contribute significantly to spin-dependent 
	  scattering. Because their orbits extend further
	  out than those in the core, the form factor near
	  $q=0$, which reflects the mean square radius of 
	  the contributing nucleons, will always decrease
	  faster for spin-dependent scattering
	  than for spin-independent scattering.
	  By the same token though, the Fermi-surface 
	  nucleons have a higher momentum on the average, 
	  and the spin-dependent form factors will
	  therefore be larger at high $q$ than their
	  spin-independent conterpartners.
	  
	 On the other hand, for rather light isotopes (F, Na, Si) 
         Vergados et al. 
\cite{Divari:2000dc} claimed the opposite --- the drop 
         of the scalar form factor with increasing $q$ 
	 is less dramatic compared
         to that of the spin structure functions.
	 Therefore, contrary to heavy-mass targets, 
	 the light-mass targets have a
	 better sensitivity to the scalar interaction for 
	 rather large $q$ than to the spin one.
	 Nevertheless, this advantage looks illusive 
	 due to the weak $A^2$-enhancement of scalar
	 interactions for the low-mass isotopes.

\subsection{Tellurium, $^{123,125}$Te}
The Theory of Finite Fermi Systems 
	was used by Nikolaev and Klapdor-Kleingrothaus
\cite{Nikolaev:1993dd} to define the $q$-dependence of nuclear 
        form factors (structure functions) 
	in spin-dependent WIMP scattering off the $^{123}$Te isotope.
	This approach allows one to describe the reduction of 
	single-particle spin-dependent matrix elements
	in the nuclear medium.
         Unfortunately, only structure functions 
	 for nucleus spin interaction with the
	 {\em pure bino-like}\/ ($a_0/a_1 = 0.297$) 
	 lightest neutralino are presented in 
\cite{Nikolaev:1993dd}. 
         Therefore it is not possible to extract 
	 the partial structure functions
	 $S^{123}_{ij}(q)$ for the $^{123}$Te isotopes from
\cite{Nikolaev:1993dd}. 

	Ressell and Dean 
	have performed most accurate nuclear 
	calculations of the neutralino-nucleus spin-dependent
	cross section for $^{125}$Te
	(together with $^{129,131}$Xe and $^{127}$I) in 
\cite{Ressell:1997kx}.
         The details of the shell model and residual 
	 nucleon-nucleon interactions used in the calculations are given in 
	 the next Section. 
         The $q$-dependence of complete spin structure functions
 	 $S^{125}_{ij}(q)$ for $^{125}$Te 
	 one can br found in Fig.~2 of 
\cite{Ressell:1997kx}. 
	Ressell and Dean
	gave rather simple parameterizations of the 
	complete structure functions $S^{125}_{ij}(q)$ as tables
	of the coefficients $C_k$ 
(given in Table~\ref{R-D:I-125:6th}) of 6th order polynomials in $y$:
$\displaystyle S^{125}_{ij}(q) = \sum_{k = 0}^6 C_k y^k$,   
	where $y = (qb/2)^2$ and $b=2.24$~fm~$=11.35/$GeV for $^{125}$Te.
	These so-called abbreviated structure functions
	are only valid for $y \leq 1$ (or $q= 2\sqrt{y}/b \leq 176$~MeV)
	and therefore for the recoil energy $E \le 124$~keV.
\begin{table}[h!]
\caption{~~$^{125}$Te isotope.
  The first column gives the order of $y^k$, the next
  three columns give the corresponding values of the $C_k$ for $S^{125}_{00}$,
  $S^{125}_{01}$, and $S^{125}_{11}$ for the Bonn A calculation.  The last
  three columns present the same results for the Nijmegen II calculation.
From \protect\cite{Ressell:1997kx}.}
\begin{tabular}{|r|r|r|r||r|r|r|} \hline
&\multicolumn{3}{|c||}{Bonn A} &\multicolumn{3}{|c|}{Nijmegen II} \\ \hline
&~~~~$S^{125}_{00}$~~~~&~~~~$S^{125}_{01}$~~~~&~~~~$S^{125}_{11}$~~~~  
&~~~~$S^{125}_{00}$~~~~&~~~~$S^{125}_{01}$~~~~&~~~~$S^{125}_{11}$~~~~ \\ \hline
$y^0$ & 0.0397   &$-0.0789$ & 0.0392  & 0.0496  & $-0.0993$& 0.0496 \\
$y^1$ & $-0.2712$&  0.5727  &$-0.3004$&$-0.3425$&  0.7315  &$-0.3875$ \\
$y^2$ &   0.8694 &$-1.9007$ &  1.0378 &  1.0666 &$-2.3930$ &  1.3290 \\
$y^3$ & $-1.5695$&  3.4698  &$-1.9460$&$-1.8547$&  4.3229  &$-2.4952$ \\
$y^4$ &   1.6184 &$-3.5546$ &  2.0264 &  1.8464 &$-4.4282$ &  2.6334 \\
$y^5$ & $-0.8797$&  1.9020  &$-1.0944$&$-0.9762$&  2.3905  &$-1.4535$ \\
$y^6$ &   0.1941 &$-0.4116$ &  0.2380 &  0.2110 &$-0.5245$ &  0.3241 \\
\hline
\end{tabular}
\label{R-D:I-125:6th} 
\bigskip 
\caption{~~$^{125}$Te isotope.
The first column gives the order of $y^k$, the next three
columns give the corresponding values of the $C_k$ for $S^{125}_{00}$,
$S^{125}_{01}$, and $S^{125}_{11}$ for the Bonn A calculation.  The last
three columns present the same results for the Nijmegen II calculation.  
From \protect\cite{Ressell:1997kx}.}
\begin{tabular}{|r|r|r|r||r|r|r|} \hline
&\multicolumn{3}{|c||}{Bonn A}&\multicolumn{3}{|c|}{Nijmegen II} \\ \hline
$\times (e^{-2 y})$ 
&~~~~~$S^{125}_{00}$~~~~&~~~~~$S^{125}_{01}$~~~~&~~~~~$S^{125}_{11}$~~~~  
&~~~~~$S^{125}_{00}$~~~~&~~~~~$S^{125}_{01}$~~~~&~~~~~$S^{125}_{11}$~~~~ \\ \hline
$y^0$&  0.03971 &$-0.07894$&  0.03922 &  0.04960 &$-0.09939$&$-1.92941$\\ 
$y^1$&$-0.19610$&  0.42738 &$-0.22938$&$-0.24777$&  0.54303 &  1.68075 \\
$y^2$&  0.47265 &$-1.09331$&  0.62215 &  0.54766 &$-1.28816$&$-1.16336$\\
$y^3$&$-0.65023$&  1.55324 &$-0.92253$&$-0.66553$&  1.67206 &  0.58650 \\
$y^4$&  0.54193 &$-1.28933$&  0.78465 &  0.47462 &$-1.26883$&$-0.20730$\\
$y^5$&$-0.26456$&  0.61844 &$-0.38245$&$-0.19944$&  0.56728 &  0.05141 \\
$y^6$&  0.07489 &$-0.16964$&  0.10571 &  0.04819 &$-0.14545$&$-0.00870$\\
$y^7$&$-0.01146$&  0.02481 &$-0.01542$&$-0.00616$&  0.01959 &  0.00087 \\
$y^8$&  0.00075 &$-0.00152$&  0.00093 &  0.00032 &$-0.00107$&  0.00004 \\
${1\over{1 + y}}$& 0.0  & 0.0 & 0.0 & 0.0 & 0.0 & 1.97923 \\
\hline
\end{tabular}
\label{R-D:I-125:8th} 
\end{table}

\begin{figure}[h!] 
\begin{minipage}[b]{0.55\textwidth}
{\begin{picture}(50,75)
\put(-56,-80){\includegraphics{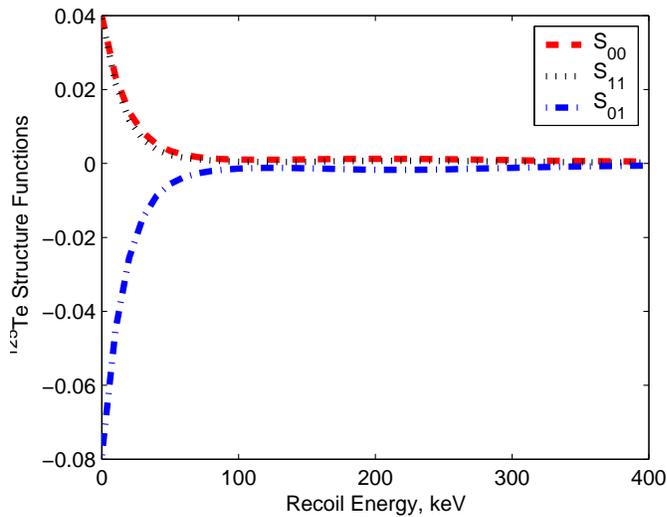}}
\end{picture}
}\end{minipage}
\hfill
\begin{minipage}[b]{0.40\textwidth}{
\caption{Partial structure functions 
        $S^{125}_{00}(q)$ (top), 
        $S^{125}_{01}(q)$ (bottom) and 
        $S^{125}_{11}(q)$ (middle) 
	in ${}^{125}$Te as a function of recoil energy  
	obtained from Ressell and Dean
\cite{Ressell:1997kx} by means of parameterization
(\ref{S125-RD-full}).
       Note that for ${}^{125}$Te, when 
       the maximal WIMP velocity $v_{\max}=600$~km/s, one has 
       $q_{\max}= 464$~MeV/$c$, $E_{\max} = 919$~keV and 
       $y_{\max}=6.95$. 
\protect\\
}
\label{S125-bva} 
}\end{minipage}
\end{figure} 
	Ressell and Dean also gave their best parameterization of 
	the complete structure functions $S^{125}_{ij}(q)$ 
	in the form of 8th order polynomials in $y$
	multiplied by a factor of $\exp(-2 y)$: 
\begin{equation}\label{S125-RD-full}
\displaystyle 
S^{125}_{ij}(q) 
= \left(\sum_{k = 0}^8 C_k y^k + C_9 {1\over{1 + y}}\right) e^{-2y}.
\end{equation}
	All coefficients $C_k$ for the $^{125}$Te isotope are given in 
Table~\ref{R-D:I-125:8th}. 
	These so-called {\em full structure functions}\/ 
	$S^{125}_{ij}(q)$ are valid for all 
	kinetically available $y \le 10$ (or $q<557$~MeV)
	and are presented in 
Fig.~\ref{S125-bva} for the Born A potential (three left columns in 
Table~\ref{R-D:I-125:8th}).

\subsection{Iodine, $^{127}$I}
\label{Iodine}
	The iodine isotope $^{127}$I is a decisive component  
	of large sodium iodide (NaI) crystals and 
	therefore plays the most 
	important role in the modern dark matter search
	(see for example DAMA 
\cite{Bernabei:2003ky,Bernabei:2003za}
         and ELEGANT, NAIAD and ANAIS experiments 
\cite{Ahmed:2003su,Cebrian:2002vd,Yoshida:2000df}). 

	Ressell and Dean 
	have performed the most accurate nuclear shell-model 
	calculations of the neutralino-nucleus spin-dependent
	cross section for $^{127}$I in 
\cite{Ressell:1997kx}.
        Within the framework of their approach 
	Ressell and Dean considered two residual nuclear interactions 
	based upon recently developed realistic nucleon-nucleon Bonn A 
\cite{Hjorth-Jensen:1995ap} and Nijmegen II 
\cite{Stoks:1994wp} potentials.  
	These two nucleon-nucleon potentials were used 
	in order to investigate the sensitivity of the results 
	to the particular nuclear Hamiltonian.
	The Bonn-A-based Hamiltonian was derived for the model 
	space consisting of the
	$1g_{7/2}, \, 2d_{5/2}, \, 3s_{1/2}, \, 2d_{3/2},$ and
	$1h_{11/2}$ orbitals, 
	allowing one to include all relevant correlations.
	In order to get good agreement
	with observables for nuclei with $A \approx 130$, 
	the single-particle energies (SPEs) were adjusted.  
	The SPEs were varied 
	until reasonable agreement between calculation
	and experiment was found for the magnetic moment, 
	the low-lying excited state energy
	spectrum, and the quadrupole moment 
	of $^{127}$I.
	The similar procedure was used by Ressell et al. in 
\cite{Ressell:1993qm}.	
  	Once the SPEs are specified, 
	a reasonable Hamiltonian can be used for the nuclei 
	under investigation 
($^{127}$I, $^{129,131}$Xe and $^{125}$Te).
	The same scheme was used for 
	the Nijmegen II-based Hamiltonian.

	To perform a full basis calculation of the
	$^{127}$I ground state properties in the space consisting of the
	$1g_{7/2}, \, 2d_{5/2}, \, 3s_{1/2}, \, 2d_{3/2},$ and
	$1h_{11/2}$ orbitals, one would need basis states consisting
	of roughly $1.3 \times 10^9$ Slater Determinants (SDs).
	Current calculations 
	can diagonalize matrices with basis dimensions in the range
	1--2$\times 10^7$ SDs.
	Therefore clearly severe truncations of the model space are needed
\cite{Ressell:1997kx}.  
	Fortunately, given the size of the model
	spaces that can be treated, a truncation scheme that includes
	the majority of relevant configurations can be devised.  
  	Finally (after relevant truncations, see 
\cite{Ressell:1997kx} for details) the m-scheme dimension of
	the $^{127}$I  model space is about 3 million SDs.
	The calculated observables agree well with experiment.
	These interactions do not seem to prefer
	excitation of more than one extra neutron pair to 
	the $1h_{11/2}$.  
	Most configurations have six neutrons in that orbital, 
	while eight are allowed.  
	Hence, this model space is more than adequate.
	It is this truncation scheme that was also 
	used for the two xenon isotopes considered ($A = 129$ and $131$). 

	In almost every instance, the results of Ressell and Dean 
\cite{Ressell:1997kx}
(Tables~8--10 of 
\cite{Bednyakov:2004xq})
	show that the spin
	$\langle {\bf S}_i \rangle $ ($i = p,n$) carried by the unpaired
	nucleon is greater than that found in the other nuclear
	models. 
	Despite these larger values for $\langle {\bf S}_i \rangle $,
	these calculations have 
	the magnetic moment 
	in good agreement with experiment in all cases.
	The larger values of $\langle {\bf S}_i \rangle $ 
	are due to the fact that more excitations of the
	even group of the nuclei were allowed
\cite{Ressell:1997kx}.
	There are visible 
	differences in the response due to the two different forces.
	In all cases reasonable
	agreement between calculation and experiment for the magnetic
	moment (using free particle $g$-factors) is achieved.
	It is obvious 
	that the differences between the two calculations
	are non-trivial but they are quite a bit smaller
	than the differences coming from the use of alternative 
	nuclear models.  
	This shows that the interaction is not the primary
	uncertainty in calculations of the 
	neutralino-nucleus spin cross sections
\cite{Ressell:1997kx}.
	The results obtained by Ressell and Dean give a factor 
	of $20$ increase in iodine's sensitivity to spin-dependent
	scattering over that previously assumed.  
	Due to the form factor suppression 
	the spin response of the sodium iodide detector's 
	is still dominated by $^{23}$Na but 
	not to the extent previously thought
\cite{Ressell:1997kx}.

        The reduced matrix elements of the multipoles in
	the definition of the $S^A(q)$
(\ref{SF-definition})
	are easily evaluated in the harmonic oscillator basis in the
	nuclear shell model.  
	Almost all calculations of $S^A(q)$ used bases of these
	harmonic oscillator wave functions.  
        In 
\cite{Ressell:1997kx}
        Ressell and Dean used more realistic 
	Woods-Saxon wave functions to evaluate
(\ref{SF-definition}).
         To check influence of the wave function basis 
         the Bonn A structure function $S^{127}(q)$
	 for the $^{127}$I isotope was also calculated with 
	 the harmonic oscillator wave functions.  
	 The Woods-Saxon wave functions made a significant
	 difference at extremely high momentum transfers when compared to
	 the usual harmonic oscillator wave functions.  
         At more modest momentum transfers 
	 the difference is found to be small.
	
	The oscillator parameter,
	$b = 1\,{\rm fm}\, A^{1/6}$, is usually retained as 
	the size parameter in Woods-Saxon evaluations of $S^A(q)$. 
	In 
\cite{Ressell:1997kx}
	a standard $sd$-shell parameterization is used:
	$b = (41.467/\hbar\omega)^{1/2}$~fm with 
	$\hbar\omega =  45 A^{-1/3} - 25 A^{-2/3}$~MeV.  
        Therefore for all isotopes with $A\approx 127$
	one has $b = 2.282$~fm. 
	In general, for heavy enough nuclei (say, with $A > 100$) 
	it is especially useful to present 
	structure functions in terms of dimensionless variable 
	$y \equiv (qb/2)^2$.  
	For $y \ll 1$ ($y \ge 1$) 
	the effects of finite momentum transfers 
	are usually rather small (quite large). 
	For these nuclei $y_{\max} = (q_{\max}b/2)^2 \simeq 10 \gg 1$, 
	and nuclear structure 
	form factors are very important. 
	Nevertheless these extremely large values of $y$ are only valid 
	for extremely massive WIMPs ($m_\chi \gg m_A$)
	moving with an almost escape velocity ($v_{\max}\approx 700$~km/s).  
	A more realistic WIMP with mass of about 100 GeV$/c^2$ 
	moving at an average velocity $v \approx \langle v \rangle = 10^{-3}c$ 
	would have
	$y_{\max} \simeq 0.4$.
	Anyway, in order to cover the entire elevant 
	neutralino parameter space,
        the structure functions 
	were evaluated up to $y = 10$
\cite{Ressell:1997kx}.

       The complete partial spin functions $S^A_{ij}(y)$ 
       for $^{127}$I, $^{129,131}$Xe and $^{125}$Te 
       and $y \leq 2$ ($q^2 \simeq 60000$ MeV$^2$),
       originally calculated by Ressell and Dean with the Bonn-A-  and 
       Nijmegen-II-based Hamiltonians, are presented graphically 
       in Fig.~2 of 
\cite{Ressell:1997kx}.
       For practical use Ressell and Dean gave the simple  
       parameterization of $^{127}$I structure functions 
$\displaystyle S^{127}_{ij}(q) = \sum_{k = 0}^6 C_k y^k$
       as 6th order polynomials in $y$ with $C_k$ coefficients from 
Table~\ref{R-D:I-127:6th}.
	These so-called {\em abbreviated structure functions}\/
	are only valid for $y \leq 1$.	
	Ressell and Dean also gave so-called 
	{\em full structure functions}, which are valid for $y \le 10$
	and reproduce the complete structure functions $S^{127}(q)$ as 
	8th order polynomials in $y$ multiplied by a factor of $\exp(-2 y)$: 
\begin{equation}\label{R-D-8th:127} 
S^{127}_{ij}(q) = e^{-2y} \sum_{k = 0}^8 C_k y^k 
\end{equation} 
	All coefficients $C_k$ for the $^{127}$I isotope are given in 
Table~\ref{R-D:I-127:8th}.
         For example, with the entries 
	 in the Bonn A section of
Table~\ref{R-D:I-127:8th} one obtains the analytical forms for $^{127}$I
         spin stricture functions: 
\begin{eqnarray}\label{S127-BonnA}
S^{127}_{00}(y)&=&e^{-2y}\left(0.0983- 0.4891 y + 1.1402 y^2 - 1.4717 y^3 + 1.1717 y^4 
\right. 
\\ & & \,\,\,\quad  \nonumber 
\left. - 0.5646 y^5 + 0.1583 y^6 - 0.0239 y^7 + 0.0015 y^8\right),\\
S^{127}_{01}(y)&=& e^{-2y}\left(0.1199- 0.6184 y + 1.5089 y^2 - 2.0737 y^3+ 1.7731 y^4 
\right. \nonumber \\ & & \,\,\,\quad \nonumber 
\left. - 0.9036 y^5 + 0.2600 y^6 - 0.0387 y^7 + 0.0024 y^8\right), \\
S^{127}_{11}(y)&=& e^{-2y}\left(0.0366- 0.1950 y + 0.5049 y^2 - 0.7475 y^3+ 0.7043 y^4 
\right. \nonumber \\ & & \,\,\,\quad \nonumber 
\left. - 0.3930 y^5 + 0.1219 y^6 - 0.0192 y^7 + 0.0012 y^8\right). 
\end{eqnarray}

\begin{table}[h!] 
\caption{~~$^{127}$I isotope.
  The first column gives the order of $y^k$, the next three
  columns give the corresponding values of the $C_k$ for $S^{127}_{00}$,
  $S^{127}_{01}$, and $S^{127}_{11}$ for the Bonn A calculation.  The last
  three columns present the same results for the Nijmegen II calculation.
From \protect\cite{Ressell:1997kx}.}
\begin{tabular}{|r|r|r|r||r|r|r|} \hline
&\multicolumn{3}{|c||}{Bonn A} &\multicolumn{3}{|c|}{Nijmegen II} \\ \hline
&~~~~$S^{127}_{00}$~~~~&~~~~$S^{127}_{01}$~~~~&~~~~$S^{127}_{11}$~~~~  
&~~~~$S^{127}_{00}$~~~~&~~~~$S^{127}_{01}$~~~~&~~~~$S^{127}_{11}$~~~~ \\ \hline
$y^0$ &  0.0983 &  0.1199 &  0.0365 &  0.1165 &  0.1619 &  0.0562 \\
$y^1$ &$-0.6750$&$-0.8436$&$-0.2627$&$-0.7923$&$-1.1403$&$-0.4085$ \\
$y^2$ &  2.1353 &  2.7354 &  0.8751 &  2.4985 &  3.7144 &  1.3778 \\
$y^3$ &$-3.7595$&$-4.9303$&$-1.6146$&$-4.3831$&$-6.7158$&$-2.5702$ \\
$y^4$ &  3.7774 &  5.0581 &  1.6908 &  4.3850 &  6.8938 &  2.7087 \\
$y^5$ &$-2.0091$&$-2.7361$&$-0.9302$&$-2.3222$&$-3.7259$&$-1.4945$ \\
$y^6$ &  0.4355 &  0.6008 &  0.2069 &  0.5015 &  0.8171 &  0.3329 \\\hline
\end{tabular} \label{R-D:I-127:6th}
\bigskip
\caption{~~$^{127}$I isotope.
The first column gives the order of $y^k$, the next three
 columns give the corresponding values of the $C_k$ for $S^{127}_{00}$,
$S^{127}_{01}$, and $S^{127}_{11}$ for the Bonn A calculation.  The last
three columns present the same results for the Nijmegen II calculation.  
From \protect\cite{Ressell:1997kx}.}
\begin{tabular}{|r|r|r|r||r|r|r|} \hline
&\multicolumn{3}{|c||}{Bonn A}&\multicolumn{3}{|c|}{Nijmegen II}\\ \hline
$\times (e^{-2 y})$ 
&~~~~~$S^{127}_{00}$~~~~&~~~~~$S^{127}_{01}$~~~~&~~~~~$S^{127}_{11}$~~~~  
&~~~~~$S^{127}_{00}$~~~~&~~~~~$S^{127}_{01}$~~~~&~~~~~$S^{127}_{11}$~~~~ \\ \hline
$y^0$ &  0.0983 &  0.1199 &  0.0366 &  0.1166 &  0.1621 &  0.0563 \\ 
$y^1$ &$-0.4891$&$-0.6184$&$-0.1950$&$-0.5721$&$-0.8363$&$-0.3038$ \\
$y^2$ &  1.1402 &  1.5089 &  0.5049 &  1.3380 &  2.0594 &  0.7948 \\
$y^3$ &$-1.4717$&$-2.0737$&$-0.7475$&$-1.7252$&$-2.8319$&$-1.1703$ \\
$y^4$ &  1.1717 &  1.7731 &  0.7043 &  1.3774 &  2.3973 &  1.0637 \\
$y^5$ &$-0.5646$&$-0.9036$&$-0.3930$&$-0.6700$&$-1.2121$&$-0.5713$ \\
$y^6$ &  0.1583 &  0.2600 &  0.1219 &  0.1905 &  0.3486 &  0.1722 \\
$y^7$ &$-0.0239$&$-0.0387$&$-0.0192$&$-0.0292$&$-0.0522$&$-0.0266$\\
$y^8$ &  0.0015 &  0.0024 &  0.0012 &  0.0019 &  0.0032 &  0.0017 \\ \hline
\end{tabular}
\label{R-D:I-127:8th} 
\end{table}

\begin{figure}[h!] 
\begin{minipage}[b]{0.50\textwidth}
{\begin{picture}(50,70)
\put(-55,-80){\includegraphics{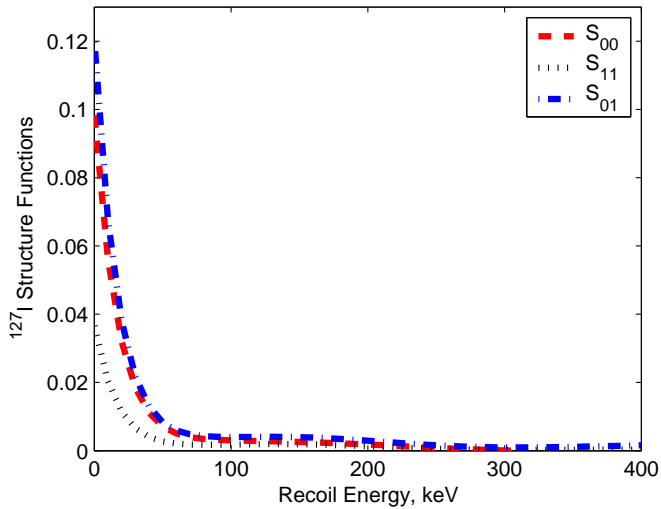}}
\end{picture}
}\end{minipage} \hfill
\begin{minipage}[b]{0.40\textwidth}{
\caption{Partial structure functions 
         of Ressell and Dean $S^{127}_{ij}$ for ${}^{127}$I as a
         function of recoil energy calculated from formulas
 (\ref{S127-BonnA}).
       Note, for ${}^{127}$I, when 
       the maximal WIMP velocity $v_{\max}=600$~km/s, one has 
       $q_{\max}\approx 472$~MeV/$c$, $E_{\max} \approx 934$~keV and 
       $y_{\max}\approx 7.44$. \protect\\
}\label{S127-bva} 
}\end{minipage}
\end{figure} 

	The structure functions $S^{127}_{ij}(y)$ for the $^{127}$I isotope 
	and for both nucleon-nucleon forces (Bonn A and Nijmegen II)
	up to $y = 10$ were carefully analyzed in 
\cite{Ressell:1997kx}. 
       Some similarities and differences between these
       two sets of iodine structure functions were observed.
       Despite that the differences between the two calculations
       are non-trivial but they are quite smaller
       than the differences coming from the use of alternative 
       nuclear models.  
       This shows, in particular,  that the interaction is not the primary
       uncertainty in calculations of the
       neutralino-nucleus scattering cross sections
\cite{Ressell:1997kx}.

Finally, it is perhaps a right place to note that 
for $y \approx y_{\max} \approx 7.5$, which already  
corresponds to a rather large energy 
($E_{\max}\approx 900$~keV) transmitted to the target nucleus, 
one may also can expect a non-negligible contribution 
from {\em inelastic}\/ WIMP-nuclear interactions not considered here.

\subsection{Xenon, $^{131}$Xe and $^{129}$Xe}
\begin{figure}[h!] 
\begin{picture}(50,70)
\put(-93,120){\includegraphics{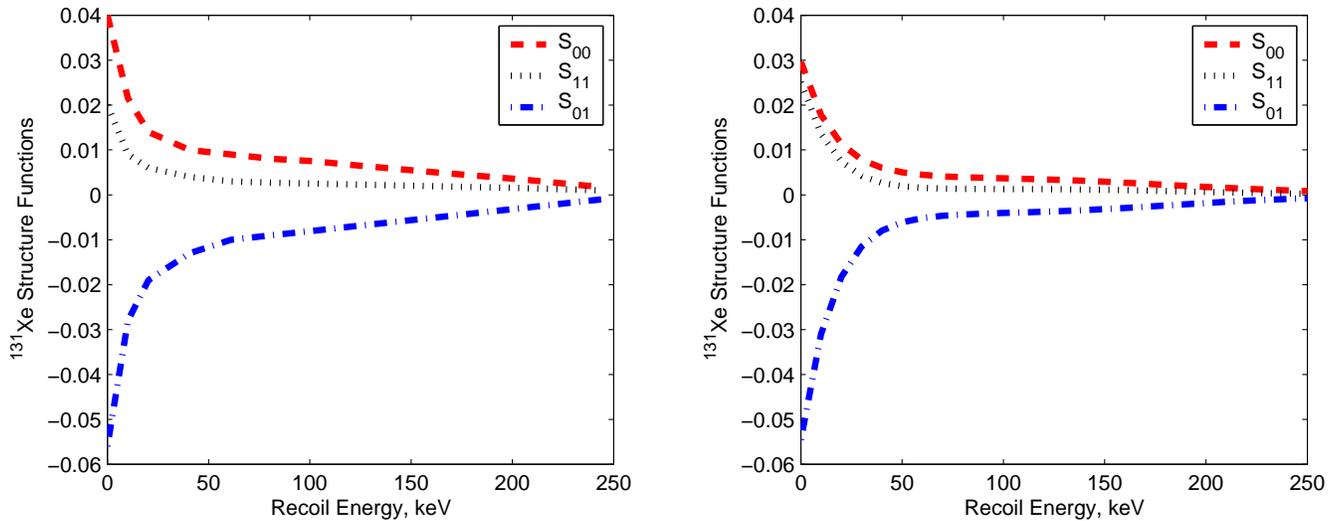}}
\end{picture}
\caption{Partial structure functions 
        $S^{131}_{00}(q)$ (top), 
        $S^{131}_{01}(q)$ (bottom) and 
        $S^{131}_{11}(q)$ (middle) 
	in ${}^{131}$Xe as a function of the recoil energy. 
	Left: results of 
\cite{Engel:1991wq} from 
Table~\ref{S131-digitaized}.
       Right: the full parameterizations of 
 \cite{Ressell:1997kx}
       for the Bonn A potential (three left columns in
Table~\ref{R-D:I-129-131:8th}).
       Note, that for ${}^{131}$Xe, when 
       the maximal WIMP velocity $v_{\max}=600$~km/s, one has 
       $q_{\max}\approx 487$~MeV/$c$, $E_{\max} \approx 963$~keV and 
       $y_{\max}\approx 7.92$. 
}\label{S131-bva} 
\end{figure}

         Xenon (gaseous and liquid) is a very popular 
	 target material for modern
	 large-scale dark matter detectors
(see, for example, 
\cite{Hart:2002qh,Luscher:2003er,Bernabei:2002qg}). 
         For the first time spin-dependent scattering
	 of SUSY-like dark matter particles from 
	 $^{131}$Xe ($J=3/2$) nuclei at non-zero momentum transfer 
	 was considered by Engel in 
\cite{Engel:1991wq}.
	The configuration-mixing
	quasiparticle Tamm-Dancoff approximation ({QTDA}) was used. 
	In the zeroth order the ground state of $^{131}$Xe was represented
	as the $1d_{3/2}$ quasineutron excitation of 
	the even-even core $|0\rangle$ treated in the BCS approximation 
	(BCS-based model of the Fermi surface).
	In order to incorporate nuclear structure corrections 
	originating from the residual interaction 
	three-quasiparticle configurations of the form 
$[\nu^{\dag}_{d3/2} [\nu^{\dag}_k\nu^{\dag}_l]^K ]^{3/2} |0\rangle$ and 
$[\nu^{\dag}_{d3/2} [\pi^{\dag}_k\pi^{\dag}_l]^K ]^{3/2} |0\rangle$ 
	were admixed. 
	Here $\pi^{\dag}$ and $\nu^{\dag}$ represent the proton and 
	neutron quasiparticle creation operators, $K$ is an arbitrary
	intermediate angular momentum, and $k,l$ run over
	a valence space consisting of the $2s$, $1d$, $0g$ and $0h$ 
	harmonic oscillator levels
\cite{Engel:1991wq}. 
	Despite the fact that the amplitudes associated with the 
        admixed three-quasiparticle states are small (less than 5\%),
	these admixtures can lead to a substantial effect.
	The experimental value of the  magnetic moment of $^{131}$Xe, which
	is about $0.69\, \mu_N$, 
	was reproduced with an accuracy of 2\%.
	The same approximation scheme results in 
	    $\langle{\bf S}^{131}_p\rangle = -0.041$ 
	and $\langle{\bf S}^{131}_n\rangle = -0.236$.
Figure~\ref{S131-bva} (left panel) 
        shows Engel's partial structure functions 
        $S^{131}_{ij}$ in ${}^{131}$Xe as a function of the recoil energy, 
	recalculated from 
Table~\ref{S131-digitaized}.
\begin{table}[h!]
\caption{Tabulated spin structure functions $S^{131}_{ij}(q)$ from 
Fig.~3 of \protect\cite{Engel:1991wq}. 
}
\label{S131-digitaized}
\begin{tabular}{|r|r|r|r|r|r|r|r|r|r|r|r|} \hline 
$q^2$ 
&~~0	&~~0.0025& 0.005 & 0.01 &  0.015&  0.02&  0.025&  0.03&   0.04&   0.05&   0.06\\
\hline \vphantom{\Huge Y}
$S^{131}_{00}(q)$ 
&~~0.4 &0.0215& 0.014&  0.01&   0.009& 0.008 & 0.0075& 0.0066 & 0.005 & 0.0035&0.0017 \\
$S^{131}_{11}(q)$ 
&0.020 & 0.009 & 0.006&  0.004 & 0.003 &0.0027 &0.0025& 0.0023 &0.0019 &0.0015&0.001 \\
$S^{131}_{01}(q)$ 
&$-0.056$&~$-0.028$&~$-0.019$&~$-0.013$&~$-0.01$&~$-0.009$&~$-0.008$&~$-0.007$
&~$-0.005$&~$-0.003$&$-0.001$\\\hline
\end{tabular}
\end{table}

        Engel noted again that the spin-dependent cross section in $^{131}$Xe 
	falls as the momentum transfer increases, but {\em more slowly}\/
	than the spin-independent cross section. 
	The spin-dependent efficiency is higher than the spin-independent one, 
	being substantially higher for very heavy neutralinos. 
	The relatively long tail of the spin-dependent structure functions 
	is caused by nucleons near the Fermi surface, which
	do the bulk of the scattering. 
	The core nucleons, which dominate the spin-independent cross section, 
	contribute much less at large $q$.
	These are very general statements that should apply to other 
	heavy nuclei as well
\cite{Engel:1991wq}. 

	The Theory of Finite Fermi Systems (TFFS) was also used
 	by Nikolaev and Klapdor-Kleingrothaus
\cite{Nikolaev:1993dd} to define the $q$-dependence of nuclear 
        structure functions (form factors) 
	in spin-dependent WIMP scattering off the $^{131}$Xe
 	isotope (as well as $^{123}$Te).
	The quenching effect (at zero and relatively low $q$), 
	due to reduction of 
	single-particle spin-dependent matrix elements in the nuclear medium,
	and its disappearance at higher $q$ was observed in 
\cite{Nikolaev:1993dd}. 
        The same $q$-behavior was also discussed before by Engel at al.
\cite{Engel:1991wq,Engel:1992qb}.
        Nevertheless the shape of the $S^{131}(q)$ from
\cite{Nikolaev:1993dd} 
	differs a bit (at intermediate $q$) 
	from the one obtained with the oscillator basis in 
\cite{Engel:1991wq}.
         Unfortunately, 
         partial structure functions
	 $S^{131}_{ij}(q)$ for $^{131}$Xe are not available from 
\cite{Nikolaev:1993dd}. 

         The most sophisticated shell-model treatment of the 
	 spin structure functions in 
	 $^{131}$Xe and $^{129}$Xe (as well as in $^{127}$I and 
	 $^{125}$Te isotopes) was performed by  Ressell and Dean in 
\cite{Ressell:1997kx}.
         The details of the calculations are given in 
	 the previous Section. 
\begin{table}[b!] 
\caption{~~$^{129,131}$Xe isotopes.
  The first column gives the order of $y^k$, the next
  three columns give the corresponding values of the $C_k$ for 
  $S^{129,131}_{00}$,
  $S^{129,131}_{01}$, and $S^{129,131}_{11}$ for the Bonn A
  calculation.  
  The last three
  columns present the same results for the Nijmegen II calculation.
From \protect\cite{Ressell:1997kx}.}
\label{R-D:I-129-131:6th}
\begin{tabular}{|r|r|r|r||r|r|r|} \hline
&\multicolumn{3}{|c||}{Bonn A} &\multicolumn{3}{|c|}{Nijmegen II} \\ \hline
&~~~~$S^{129}_{00}$~~~~&~~~~$S^{129}_{01}$~~~~&~~~~$S^{129}_{11}$~~~~  
&~~~~$S^{129}_{00}$~~~~&~~~~$S^{129}_{01}$~~~~&~~~~$S^{129}_{11}$~~~~ \\ \hline
$y^0$ &  0.0713 &$-0.1216$&  0.0518 &  0.0465 &$-0.0853$&  0.0392 \\
$y^1$ &$-0.4804$&  0.8745 &$-0.3949$&$-0.3138$&  0.6150 &$-0.2991$ \\
$y^2$ &  1.4726 &$-2.8317$&  1.3433 &  0.9656 &$-1.9847$&  1.0087 \\
$y^3$ &$-2.5323$&  5.0922 &$-2.5152$&$-1.6666$&  3.5496 &$-1.8648$ \\
$y^4$ &  2.4968 &$-5.1976$&  2.6480 &  1.6477 &$-3.6023$&  1.9399 \\
$y^5$ &$-1.3071$&  2.7924 &$-1.4557$&$-0.8642$&  1.9257 &$-1.0564$ \\
$y^6$ &  0.2796 &$-0.6088$&  0.3228 &  0.1851 &$-0.4182$&  0.2326 \\
\hline 
&~~~~$S^{131}_{00}$~~~~&~~~~$S^{131}_{01}$~~~~&~~~~$S^{131}_{11}$~~~~  
&~~~~$S^{131}_{00}$~~~~&~~~~$S^{131}_{01}$~~~~&~~~~$S^{131}_{11}$~~~~ \\ \hline
$y^0$ &  0.0296 &$-0.0545$&  0.0251 &  0.0277 &$-0.0497$&  0.0223 \\
$y^1$ &$-0.1852$&  0.3676 &$-0.1812$&$-0.1754$&  0.3389 &$-0.1627$ \\
$y^2$ &  0.5934 &$-1.1813$&  0.5932 &  0.5604 &$-1.1002$&  0.5427 \\
$y^3$ &$-1.0351$&  2.0529 &$-1.0389$&$-0.9969$&  1.9709 &$-0.9892$ \\
$y^4$ &  1.0049 &$-1.9827$&  1.0071 &  1.0100 &$-1.9996$&  1.0150 \\
$y^5$ &$-0.5078$&  0.9967 &$-0.5071$&$-0.5402$&  1.0681 &$-0.5459$ \\
$y^6$ &  0.1037 &$-0.2026$&  0.1031 &  0.1174 &$-0.2316$&  0.1189 \\ \hline
\end{tabular}
\end{table}
	Their complete sets of the structure functions $S^{129,131}(q)$
	are valid for all relevant values of the momentum transfer
(see Fig.~3 in 
\cite{Ressell:1997kx}). 
	Ressell and Dean 
	gave the simple parameterizations of 
	the complete structure functions $S^{129,131}(q)$ as 
	6th order polynomials in $y$ with coefficients $C_k$ in 
Table~\ref{R-D:I-129-131:6th}: 
$\displaystyle S^{A}_{ij}(q) = \sum_{k = 0}^6 C_k y^k$.  
	These so-called {\em abbreviated structure functions}\/
	are only valid for $y \leq 1$.	

	For the $^{129,131}$Xe isotopes (as previously 
	for $^{127}$I and $^{125}$Te)  
	Ressell and Dean also presented their parameterization of the  
	{\em full structure functions}\/ in the following analytical form
\begin{equation}
\label{R-D-8th}
S^A_{ij}(q) = \left(\sum_{k = 0}^8 C_k y^k 
+ C_9 {1\over{1 + y}}\right) e^{-2y}.
\end{equation} 
	The form is also valid for $y \le 10$.  
	The relevant coefficients $C_k$ are given in 
Table~\ref{R-D:I-129-131:8th}. 
\begin{table}[ht!] 
\caption{~~$^{129,131}$Xe isotope.
The first column gives the order of $y^k$, the next three
columns give the corresponding values of the $C_k$ for $S^{129,131}_{00}$,
$S^{129,131}_{01}$, and $S^{129,131}_{11}$ for the Bonn A calculation.  
The last
three columns present the same results for the Nijmegen II calculation.  
From \protect\cite{Ressell:1997kx}.} 
\label{R-D:I-129-131:8th}  
\begin{tabular}{|r|r|r|r||r|r|r|} \hline
&\multicolumn{3}{|c||}{Bonn A} &\multicolumn{3}{|c|}{Nijmegen II} \\ \hline
$\times e^{-2 y}$ 
&~~~~$S^{129}_{00}$~~~~&~~~~$S^{129}_{01}$~~~~&~~~~$S^{129}_{11}$~~~~  
&~~~~$S^{129}_{00}$~~~~&~~~~$S^{129}_{01}$~~~~&~~~~$S^{129}_{11}$~~~~ \\ \hline
$y^0$ &  0.07132 &$-0.12166$&$-2.05825$&  0.04649 &$-0.08538$&$-1.28214$ \\ 
$y^1$ &$-0.34478$&  0.64435 &  1.80756 &$-0.22551$&  0.45343 &  1.09276 \\
$y^2$ &  0.75590 &$-1.52732$&$-1.27746$&  0.49905 &$-1.06546$&$-0.71295$ \\
$y^3$ &$-0.93345$&  2.02061 &  0.65459 &$-0.62244$&  1.38670 &  0.31489 \\
$y^4$ &  0.69006 &$-1.57689$&$-0.22197$&  0.46361 &$-1.05940$&$-0.08351$ \\
$y^5$ &$-0.30248$&  0.72398 &  0.04546 &$-0.20375$&  0.47576 &  0.01059 \\
$y^6$ &  0.07653 &$-0.19040$&$-0.00427$&  0.05109 &$-0.12208$&  0.00023 \\
$y^7$ &$-0.01032$&  0.02638 &$-0.00014$&$-0.00671$&  0.01643 &$-0.00024$ \\
$y^8$ &  0.00057 &$-0.00149$&  0.00004 &  0.00036 &$-0.00089$&  0.00002 \\
${1\over{1 + y}}$ & 0.0 & 0.0 & 2.11016 & 0.0 & 0.0 & 1.32136 \\ \hline
$\times e^{-2 y}$ 
&~~~~$S^{131}_{00}$~~~~&~~~~$S^{131}_{01}$~~~~&~~~~$S^{131}_{11}$~~~~  
&~~~~$S^{131}_{00}$~~~~&~~~~$S^{131}_{01}$~~~~&~~~~$S^{131}_{11}$~~~~ \\ \hline
$y^0$ &  0.02964 &$-0.05455$&  0.02510 &  0.02773 &$-0.04978$&  0.02234 \\ 
$y^1$ &$-0.13343$&  0.27176 &$-0.13772$&$-0.12449$&  0.24725 &$-0.12206$ \\
$y^2$ &  0.37799 &$-0.72302$&  0.36661 &  0.32829 &$-0.63231$&  0.31949 \\
$y^3$ &$-0.57961$&  1.05450 &$-0.53851$&$-0.48140$&  0.89642 &$-0.46695$ \\
$y^4$ &  0.57890 &$-0.97133$&  0.49255 &  0.47565 &$-0.81645$&  0.42877 \\
$y^5$ &$-0.34556$&  0.53842 &$-0.26990$&$-0.28518$&  0.45235 &$-0.23679$ \\
$y^6$ &  0.11595 &$-0.16899$&  0.08369 &  0.09682 &$-0.14267$&  0.07408 \\
$y^7$ &$-0.02012$&  0.02742 &$-0.01340$&$-0.01710$&  0.02335 &$-0.01197$ \\
$y^8$ &  0.00142 &$-0.00181$&  0.00087 &  0.00124 &$-0.00156$&  0.00079 \\\hline
\hline\end{tabular} 
\end{table}
Figure~\ref{S131-bva} (right panel)
	presents parameterizations $S^{131}_{ij}$ from (\ref{R-D-8th})
       for the Bonn A potential (three left columns in
Table~\ref{R-D:I-129-131:8th}) 
	as a function of the recoil energy.
\begin{figure}[h!] 
\begin{picture}(50,70)
\put(-93,120){\includegraphics{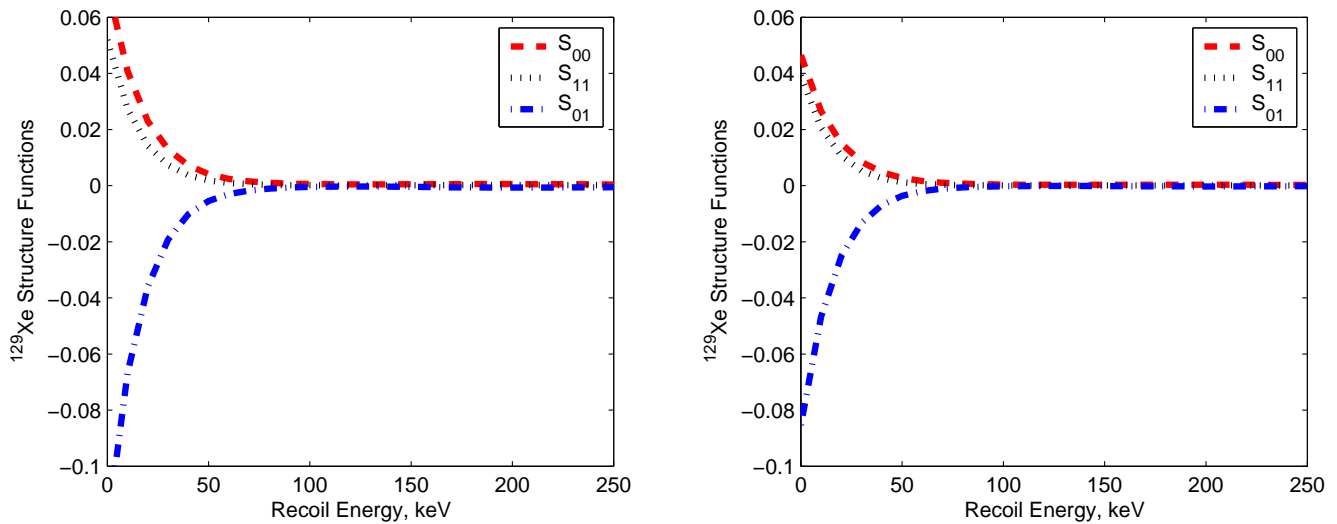}}
\end{picture}
\caption{Partial structure functions 
        $S^{129}_{00}(q)$ (top), 
        $S^{129}_{01}(q)$ (bottom) and 
        $S^{129}_{11}(q)$ (middle) 
	in ${}^{129}$Xe from 
 \cite{Ressell:1997kx} as a function of the recoil energy. 
	Left: the full parameterizations 
       for the Bonn A potential (three left columns in
Table~\ref{R-D:I-129-131:8th}).
       Right: the full parameterizations 
       for the Nijmegen II potential (three right columns in
Table~\ref{R-D:I-129-131:8th}).
}\label{S129-bva} 
\end{figure} 
      From
Fig.~\ref{S129-bva}, where parameterizations of the 
      full structure functions $S^{129}_{ij}$ are depicted
(see Table~\ref{R-D:I-129-131:8th} for ${}^{129}$Xe),  
      one can make a general conclusion 
      that there is no large difference
      between the calculations performed in 
\cite{Ressell:1997kx} with the Bonn-A-based and 
      Nijmegen-II-based Hamiltonians.

	Structure functions 
	$S^{131}(q)$ for $^{131}$Xe have been calculated in the context of two 
	other nuclear models mentioned above, the QTDA by Engel  
\cite{Engel:1991wq} and the TFFS by Nikolaev and Klapdor-Kleingrothaus
\cite{Nikolaev:1993dd}.
        Following Ressell and Dean
\cite{Ressell:1997kx}, we briefly touch upon the 
        differences in $S^{131}(q)$ that are the result of 
	using different nuclear models.  
	All three calculations show significant quenching compared to
	the single-particle estimate.  
	The spin distributions obtained in 
	the QTDA and TFFS are somewhat different 
(see Table 10 from 
\cite{Bednyakov:2004xq})) while 
	the full structure functions (for bino neutralinos) 
	are quite similar.  
	While the values for $ \langle {\bf S}_{n} \rangle $
	differ very little
	between Ressell and Dean's and Engel's result in the QTDA, 
	the difference in the values of
	$S^{131}(0)$ is almost a factor of 2 between the two calculations.
	It should be noted as well that both the QTDA and the 
	TFFS calculations of $S^{131}(q)$ 
	asymptotically reach the single-particle structure function.  
	This is not the case in Ressell and Dean's calculations, 
	which are well below the single-particle estimate 
	for all values of $q^2$.  
	It is apparent that the shell-model-derived 
	structure functions have a much steeper
	fall-off as a function of $q^2$
\cite{Ressell:1997kx}.  
         This difference can be considered as an
         estimate of the uncertainty followed from the choice of the
	 nuclear model.  

       Finally, we note that Ressell and Dean in 
\cite{Ressell:1997kx} gave 
       accurate structure function parameterizations for
       the $^{129,131}$Xe, $^{127}$I and $^{125}$Te isotopes
       which are very useful in rather precise  
       calculations of the event rates in heavy-target 
       dark matter detectors.  
       The calculations of Ressell and Dean contain
       more excitations within the model space and use
       more modern and realistic nuclear interactions than others
       in the literature, 
       with a possible exception of the QTDA calculations 
\cite{Engel:1991wq} for $^{131}$Xe
\cite{Ressell:1997kx}.

\subsection{Lead, $^{207}$Pb}
	Among the nuclei which can be considered as targets  
	for direct dark matter detection,  
	$^{207}$Pb seems a potentially rather interesting candidate. 
	The spin matrix element of this nucleus has not been evaluated
	quite accurately since one expected that WIMP-nuclear 
	spin interaction is important only with light nuclei. 
	But the spin matrix element in the light systems is quenched. 
	On the other hand, the spin matrix element of $^{207}$Pb ($J=1/2$), 
	especially the isoscalar one, does not suffer unusually large 
	quenching, as is known from the study of the magnetic moment
\cite{Kosmas:1997jm}. 
	It is believed that $^{207}$Pb has a quite simple structure, 
	its ground state can be described as a $2p_{1/2}$ neutron hole 
	outside the doubly magic (closed-shell) nucleus $^{208}$Pb.
	Due to its low angular momentum, only two multipoles 
$L =0$ and $L =2$ 
	can contribute even at large momentum transfers. 
	This is why Kosmas and Vergados in 
\cite{Kosmas:1997jm} chose $^{207}$Pb
	for investigation of momentum transfer dependence 
	of spin matrix elements in the heaviest 
	possible nuclei relevant to dark matter search. 
	In the $q=0$ limit Vergados and Kosmas gave
	the spin matrix element in the simple form 
$|{\bf J}|^2 = \Big|f^0_A \Omega_0(0) + f^1_A \Omega_1(0)\Big|^2$,
	and found that
	$\Omega_0(0) =-0.95659/\sqrt{3}$ and 
	$\Omega_1(0) = 0.83296/\sqrt{3}$
\cite{Vergados:1996hs,Kosmas:1997jm}.
        The momentum transfer dependence of the total cross section for elastic
	scattering of cold dark matter candidates,
	i.e. the lightest supersymmetric particle (LSP), 
	from $^{207}$Pb was examined by Kosmas and Vergados in 
\cite{Kosmas:1997jm,Vergados:1996jp,Vergados:1996jn,Kosmas:1995md}.

        If the spin-dependent differential cross section is taken 
	in the form of Engel, Ressell et al.
\cite{Engel:1991wq,Ressell:1997kx,Ressell:1993qm}
	and the spin structure function $h(q)$ has the form of Vergados et al.
\cite{Divari:2000dc}:
$$ 
d\sigma = \frac{8 G^2_{\rm F}}{2J+1} h(q) dq^2, \qquad
h(q) =\frac14 \left[
                (f^0_A)^2 \, S^{207}_{00}(q)
              + (f^1_A)^2 \, S^{207}_{11}(q)
              + f^0_A \, f^1_A \, S^{207}_{01}(q)
              \right], 
$$
       then 
       the partial structure functions for $^{207}$Pb 
       $S^{207}_{00}(q)$, 
       $S^{207}_{11}(q)$, and 
       $S^{207}_{01}(q)$ have the form 
\cite{Divari:2000dc}
\begin{eqnarray}\nonumber
S^{207}_{00}(q) &=& \frac{2J+1}{16\pi} \, (0.305)\, I^{}_{00}(q),
\\ \label{SF-Pb207}
S^{207}_{11}(q) &=& \frac{2J+1}{16\pi} \, (0.231) \, I^{}_{11}(q), 
\\ \nonumber
S^{207}_{01}(q) &=& \frac{2J+1}{8\pi}\, (-0.266)\, I^{}_{01}(q).
\end{eqnarray}
      The spin-dependent neutralino-nucleon couplings $f^{0,1}_A$ are 
      analogous to the parameters $a_{0,1}$  from 
(\ref{Definitions.spin.decomposition}).
      Finite momentum dependence of these structure functions is 
      concentrated in $I^{}_{ij}(q)$, which can be defined, following 
\cite{Divari:2000dc,Kosmas:1997jm},
      as an integral over the forward scattering angle 
      $\xi = {\bf {\hat p}}_i \cdot {\bf {\hat q}} \ge 0$  
\begin{equation} 
\label{FormFactors_ij}
I^{}_{ij}(q) 
=    2 \int_0^1 \xi \, \,d\xi
     \frac{\Omega_i (q^2\xi^2)}{\Omega_i (0)} \,
     \frac{\Omega_j (q^2\xi^2)}{\Omega_j (0)} 
\end{equation} 
$${\rm with} \qquad 
\Omega_i({\bf q}) = (2J+1)^{-{1}/{2}} 
                    \langle J|| 
                    \sum_{j=1}^A {\bf \sigma}(j) 
		    {\bf \omega}_i (j)   
		    e^{-i{\bf q} \cdot {\bf x}_j }
                    |J\rangle
\quad{\rm and} \quad  
{\bf \omega}_0 (j) = 1, \quad 
{\bf \omega}_1 (j) = {\bf \tau}_3 (j).
$$
     Here ${\bf \sigma} (j)$, ${\bf \tau}_3 (j)$, ${\bf x}_j$ are 
     the spin, the third component of the isospin
     ($\tau_3 |p\rangle = |p\rangle$), and the coordinate of 
     the $j$th nucleon, ${\bf q}$ is the momentum 
     transferred to the nucleus.

     To a good approximation, the ground state of the 
      $^{207}_{~82}$Pb nucleus can be described as a $2p_{1/2}$ neutron hole 
     in the $^{208}_{~82}$Pb closed shell
\cite{Kosmas:1997jm}. 
      Then for the $L=0$ 
      multipole one finds
$$
\Omega_1({\bf q})=(1/\sqrt{3}) F_{2p} ({\bf q}^2) = - \Omega_0({\bf q}) 
\quad {\rm and} \quad 
I_{00} = I_{01} = I_{11} = 2 \int_0^1 \xi \, [ F_{2p} (q^2) ]^2 \,d\xi
= [ F_{2p} (q^2) ]^2.
$$
         Here the form factor of a single-particle harmonic
	 oscillator wave function has the form
\begin{equation}
\label{F-oscillator}
F_{nl} (q^2) = 
e^{-u/2}\sum_{\lambda=0}^{N_{\max}}\gamma_\lambda^{(nl)}(2u)^{\lambda}
\end{equation}
        with the dimensionless variable $u = q^2b^2/2$ 
	($b=2.434$~fm for Pb) and the 
        coefficients $\gamma_\lambda^{(nl)}$ given in 
Table~\ref{Gammas}.
\begin{table}[h!] 
\caption{The coefficients $\gamma_{\lambda}^{(nl)}$, 
         entering into the polynomial describing the form factor 
(\ref{F-oscillator}) of a single particle harmonic
	 oscillator wave function up to  $6 \hbar \omega$. 
	 From \cite{Kosmas:1995md}.}
\label{Gammas}
\begin{center}
\begin{tabular}{|c|ccccccc|}
\hline
$n$ $l$&~~$\lambda =0$&~~$\lambda =1$~~&~~$\lambda =2$~~&~~~~$\lambda =3$~~~~
&~~~~$\lambda =4$~~~~&~~~~$\lambda =5$~~~~&~~~~$\lambda =6$ \\ \hline
 0 0 &  1 &  &  &  &  &  &\\
 0 1 &  1 & $-1/6$ &  &  &  &$        $& \\
 1 0 &  1 & $-1/3$ & 1/24   &$        $&  &  &\\
 0 2 &  1 & $-1/3$ & 1/60   &$        $&   & & \\
 1 1 &  1 & $-1/2$ & 11/120 &$-1/240  $&  &  &\\
 0 3 &  1 & $-1/2$ & 1/20   &$-1/840  $&  &  &\\
 2 0 &  1 & $-2/3$ & 11/60  &$-1/60   $& 1/1920  &  &\\
 1 2 &  1 & $-2/3$ & 19/120 &$-11/840 $& 1/3360  &  & \\
 0 4 &  1 & $-2/3$ & 1/10   &$-1/210  $& 1/15120 &  &\\
 2 1 &  1 & $-5/6$ & 17/60  &$-31/840 $& 9/4480  &$-1/26880 $&\\
 1 3 &  1 & $-5/6$ & 29/120 &$-47/1680$& 37/30240&$-1/60480 $&\\
 0 5 &  1 & $-5/6$ & 1/6    &$-1/84   $& 1/3024  &$-1/332640$&\\
 3 0 &  1 & $ -1 $ & 17/40  &$-31/420 $& 27/4480 &$-1/4480  $& 1/322560 \\
 2 2 &  1 & $ -1 $ & 2/5    &$-1/15   $& 41/8064 &$-1/5760  $& 1/483840 \\
 1 4 &  1 & $ -1 $ & 41/120 &$-1/20   $& 1/315   &$-1/11880 $& 1/1330560 \\
 0 6 &  1 & $ -1 $ & 1/4    &$-1/42   $& 1/1008  &$-1/55440 $& 1/8648640 \\
\hline
\end{tabular}
\end{center}
\end{table}
       With entries ($n,l=2,1$) of 
Table~\ref{Gammas} the form factor squared 
$$
F^2_{2p} (q^2) = e^{-u}
\left(1-\frac{5}{3}\,u +\frac{17}{15}\,u^2-\frac{31}{105}\,u^3 
     + \frac{9}{280}\,u^{4}-\frac{1}{840}\ u^5 \right)^2,
$$
       defines the partial spin structure functions 
       for $^{207}$Pb in the simplest case (pure $2p_{1/2}$ neutron 
	 hole in the closed  shell and $L=0$). 
 
       Even though the probability of finding a pure $2p_{1/2}$ neutron hole
       in the $\frac{1}{2}^-$ ground state of  $^{207}$Pb 
       is greater than 95\%, the ground state magnetic moment 
       is quenched due to the $1^+$ $p$-$h$ excitation involving 
       the spin orbit partners. 
       Hence, 
       Kosmas and Vergados expected a similar suppression
       of the isovector spin matrix elements
\cite{Kosmas:1997jm}. 
       In this case for the $L =0$ 
       multipole one has 
$$
\Omega_{j=0,1}({\bf q}) = (-1)^{j+1}
                       C_0^2 \, \{ F_{2p} (q^2) /\sqrt{3}
                       -8  \,[  (7/13)^{1/2} C_1 F_{0i} (q^2) \,
                       + (-1)^{j} 
		\, (5/11)^{1/2} C_2 F_{0h} (q^2) ]\, \}.
$$ 
      With $C_0 = 0.973350$, $C_1 = 0.005295$, $C_2 = -0.006984$, definitions
(\ref{FormFactors_ij}) and 
(\ref{F-oscillator}) one can 
      calculate partial structure functions again.
      The situation with the multipole $L =2$ 
      contribution is more complicated, this contribution 
      appears to be rather significant especially for the large 
      momentum transfer (the details can be found in 
\cite{Kosmas:1997jm}). 
\begin{figure}[h!] 
\begin{minipage}[b]{0.55\textwidth}
{\begin{picture}(50,70)
\put(-50,-85){\includegraphics{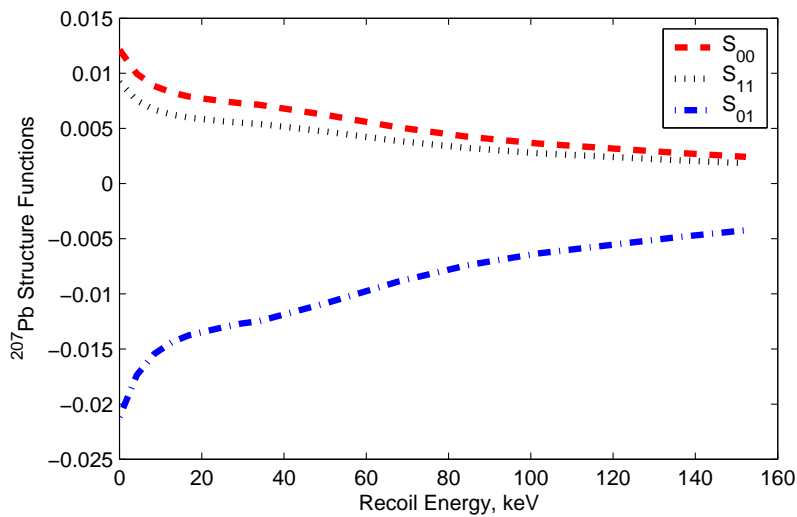}}
\end{picture}
}\end{minipage} \hfill
\begin{minipage}[b]{0.35\textwidth}{
\caption{Recoil energy dependence of the complete 
	$^{207}$Pb partial structure functions $S^{207}_{ij}$ from 
        (\ref{SF-Pb207}) of Kosmas and Vergados    
\cite{Kosmas:1997jm}. 
}\label{S207-bva} 
}\end{minipage}
\end{figure} 
	The complete 
	$^{207}$Pb partial structure functions $S^{207}_{ij}$ from 
        (\ref{SF-Pb207}) of Kosmas and Vergados    
\cite{Kosmas:1997jm} is given 
in Fig.~\ref{S207-bva} as a function of the recoil energy.

	Concluding the 
	discussion of lead structure functions we note, following 
\cite{Kosmas:1997jm},
        that for a heavy nucleus and a high WIMP mass the momentum transfer 
	dependence of the spin monopole ($L =0$) 
	matrix elements is quite large. 
	It is, however, to a large extent neutralized by the spin quadrupole 
	$(L=2)$. 
	Thus 
	the overall effect is not dramatic for WIMP masses less than 100 GeV. 
	For the spin-induced cross section of heavy 
	odd-A nuclear targets, as in the case of $^{207}$Pb, 
	the reduction is less pronounced
        since the high multipoles tend to enhance the cross section 
	as the momentum transfer increases (for LSP mass $<$ 200 GeV) and 
	partially cancel the momentum reduction 
\cite{Kosmas:1997jm}.

\section{On data analysis in finite momentum transfer framework}
        To perform the 
	data analysis in the finite momentum transfer approximation 
	directly in terms of the effective spin nucleon couplings 
	$a_{0,1}$ together with the scalar WIMP-proton cross
	section $\sigma^{p}_\SI(0)$ 
	we propose to use formulas 
	schematically given below.
        Nowadays it is also rather reasonable to assume 
        $\sigma^{p}_\SI(0)\approx \sigma^{n}_\SI(0)$.
	The differential event rate 
(\ref{Definitions.diff.rate}) can be presented as follows:
\begin{eqnarray}
\label{fit-finite-q}
\frac{dR(\eth,\emx)}{d\ER}
&=& {\cal N}(\eth,\emx,\ER,m_\chi)
     \left[ \eta^{}_\SI(\ER,m_\chi) \,\sigma^{p}_\SI
           +\eta^{\prime}_\SD(\ER,m_\chi,\omega) \,  
        {a_0^2}
     \right ];
\\ \nonumber
{\cal N}(\eth,\emx,\ER,m_\chi) 
&=&  \left[ N_T \frac{c \rho_\chi}{2 m_\chi} 
      \frac{M_A}{\mu_p^2}
      \right]
      \frac{4 \mu_A^2}{\left< q^2_{\max}\right>}
      \langle \frac{v}{c} \rangle  I(\ER)
       \theta (\ER-\eth) \theta(\emx-\ER),
\\ \nonumber
\eta^{}_\SI(\ER,m_\chi)
&=& \left\{ A^2 F^2_\SI(\ER)\right\}
; \\ \nonumber
\eta^{\prime}_\SD(\ER,m_\chi,\omega)
&=& \mu_p^2
\left\{ \frac{4 
}{2J+1} 
	\left(S_{00}(q) + \omega^2\, S_{11}(q) + \omega\, S_{01}(q) 
\right) \right\}; \\  \nonumber
I(\ER)&=& \int_{0}^{\infty}\frac{\langle v^2 \rangle}{\langle v \rangle} 
            \frac{ dv}{v} f(v)\theta(4\mu_A^2 v^2-2M_A\ER).
\end{eqnarray}
       Here $\left<q^2_{\max}\right> = 4\mu^2_A \left< v^2 \right>$
       where $\left< v \right>$ and $\left< v^2 \right>$ 
       are the mean and the mean squared velocity of WIMPs.
       The isovector-to-isoscalar nucleon couplings ratio 
       is $\omega = {a_1}/{a_0}$.
       In expressions
(\ref{fit-finite-q}) 
       are introduced the detector threshold recoil energy 
       $\eth$ and the maximal available recoil energy $\emx$ 
       ($\eth \le \ER \le \emx$).
       In practice, for example with
       an ionization or scintillation signal, one has to 
       take into account the quenching of the recoil energy, when 
       visible recoil energy is smaller then the real
       recoil energy transmitted by the WIMP to the target nucleus.

       Formulas 
(\ref{fit-finite-q}) allow experimental recoil 
       spectra to be directly described in terms of 
       only {\em three}\ 
\cite{Bednyakov:1994te} 
       independent parameters 
       ($\sigma^{p}_\SI$, $a^2_0$ and $\omega$)
       for any fixed WIMP mass $m_\chi$ (and any 
       neutralino composition).
       Contrary to some other possibilities 
       (see, for example, 
\cite{Tovey:2000mm}), 
       this procedure is direct and uses
       as much as possible the results of the 
       most accurate nuclear spin structure calculations.
       
\section{Conclusion} 
       There is continuous theoretical and experimental 
       interest in existence of dark matter of the Universe. 
       One of the best motivated non-baryonic dark matter candidates 
       is the neutralino, the lightest supersymmetric particle. 
       The motivation for supersymmetry arises
       naturally in modern theories of particle physics. 
       In this work we discussed the spin-dependent interaction 
       of neutralinos with odd-A nuclei. 
       The nuclear structure plays an important role in determining 
       the strength of the neutralino-nucleus cross section for this
       type of interaction and 
       therefore defines the sensitivity of dark matter detection. 
       In the limit of zero momentum transfer  
       the relevant physical quantities are 
       the proton and neutron spin averages $ \langle {\bf S}_{p(n)} \rangle $,
       which have to be evaluated within a proper nuclear model. 
       These values determine the event rate expected 
       in a direct dark matter search experiment  
       due to spin-dependent neutralino-nucleus interaction.
       In our previous paper 
\cite{Bednyakov:2004xq}
       the calculations of zero-momentum transfer 
       spin-dependent matrix elements was reviewed and 
       to our knowledge, a complete list of calculated
       spin matrix elements was presented for nuclei throughout 
       the periodic table
\footnote{In Table 4 of 
\cite{Bednyakov:2004xq} the results for $^{43}$Ca ($J=7/2$) 
\cite{Lewin:1996rx} were missed ---
          one has 
$\langle {\bf S}^{43}_{n} \rangle = 0.5$ in ISPSM 
and 
$\langle {\bf S}^{43}_{n} \rangle = 0.344$ ($\mu= -1.318 \mu_N$)
in OGM together with  
$\langle {\bf S}^{43}_{p} \rangle\equiv 0$. 
}.
       The general feature is that spin matrix elements 
       depend in general rather 
       sensitively on the details of the nuclear structure. 
       For a rather heavy WIMP (or light supersymmetric particle)
       and sufficiently heavy nuclei, the dependence of the nuclear matrix 
       elements on the momentum transfer cannot be ignored. 
       This affects the spin matrix elements. 

 	A comprehensive collection, to our knowledge,  of the spin-dependent 
	structure functions $S^A(q)$ calculated for finite momentum transfer  
	($q>0$) within different nuclear models is presented and discussed.
	These functions describe recoil energy dependence of the 
	differential event rate 
	due to spin-dependent neutralino-nucleon interaction,
	provided neutralino is a dark matter particle.
	Together with our 
	previous paper ``Nuclear spin structure in dark matter search: 
        The zero momentum transfer limit''
\cite{Bednyakov:2004xq}
        this paper completes our review of 
	the nuclear spin structure calculations involved 
	in the problem of direct dark matter search.

	Now that spin structure functions are available
	for almost all experimentally interesting nuclei 
	(and collected in this review),
	they could 
	be coherently used by all experimental groups. 
	This will make easier and more reliable comparisons between 
	results of different dark matter search experiments
	and put it on the equal footing
\cite{Ressell:1997kx}. 
        It will also allow one to reduce significantly the nuclear 
	physics systematic uncertainties in the analysis of the
	data. 

\smallskip
       This work was supported in part by 
       the VEGA Grant Agency of the Slovac Republic under contract
       No. 1/0249/03 and by the Russian Foundation for Basic Research
       (grant 06--02--04003).
       V.A.B. thanks V.A. Kuzmin (JINR) for careful reading of the 
       manuscript and his comments concerning the 
       maximal momentum transfer.
\bigskip

\section*{APPENDIX}
    For completeness, in this appendix we collect some formulas 
    which allow one to connect the WIMP-nucleon scattering 
    with the WIMP-nuclear scattering.
    We directly follow Engel et al.
\cite{Engel:1992bf,Engel:1991wq}. 
    The low-energy effective WIMP-nucleon Lagrangian is
$\displaystyle 
\label{Definitions.nucleon.L}
{\cal L}_{\rm eff} = \bar{\chi}\gamma^\mu\gamma_5\chi \cdot {\cal J}_\mu (x),  
$ 
    where ${\cal J}_\mu (x) \propto \bar{N} \gamma_\mu\gamma_5 N$ 
    is the nucleon current.
    The one-nucleon matrix element of the current at finite $q$
    takes the approximate form
\begin{equation}\label{Definitions.nucleon.J}
\langle p,s | {\cal J}_\mu (x) | p', s'\rangle 
= \bar{U}_N(p,s) \left(\frac{a_0+a_1\tau_3}{2}\gamma_\mu\gamma_5
+ \frac{m_N a_1 \tau_3}{q^2+m^2_\pi}q_\mu\gamma_5
                 \right) U_N(p',s') e^{iq^\nu x_\mu}.
\end{equation}
        Here $q_\mu=p_\mu-p'_\mu$, $U_N(p,s)$ is the 
	nucleon 4-component spinor
	and the energy transfer $q_0$ was 
	assumed to be very small. 
	In the nonrelativistic limit the time component of current 
(\ref{Definitions.nucleon.J}) 
	is proportional to $v/c \approx 10^{-3}$ and can be safely neglected. 
	For the spatial component of the current one has 
	the expression 
\begin{equation}\label{Definitions.nucleon.spartial}
\langle p,s | \vec{{\cal J}_\mu} (x) | p', s'\rangle = 
\langle s | 
\frac{a_0+a_1\tau_3}{2}{\vec{\sigma}} 
            - \frac{(\vec{\sigma}\cdot\vec{q})\, a_1\tau_3}{2(q^2+m^2_\pi)} 
	    \vec{q}| s'\rangle  
	    e^{iq^\nu x_\mu}
\end{equation}
      where $|s\rangle$ and $| s'\rangle$ are two-component spinors.    
      To obtain the cross section for scattering 
      of the WIMP from nuclei one must evaluate the matrix
      element of the nucleon current between many-nucleon states.
      In the impulse approximation the cross section is
\begin{equation}\label{cs1}
\frac{d\sigma^A}{dq^2}(v,q^2) = \frac{|{\cal M}|^2}{v^2(2J+1)\pi}, 
\qquad
  {\cal M} = \langle s | \vec{\sigma}_\chi | s'\rangle 
\int d^3 x \langle J,M | {\cal J}(\vec{x})
            | J,M'\rangle \times  e^{i \vec{q} \cdot \vec{x}},
\end{equation}
      where 
      $|{\cal M}|^2$ is summed over $s,s'M,M'$. 
      Here $J$ is
      the angular momentum of the ground state and the 
      nuclear current ${\cal J}(\vec{x})$ is given by the
      sum over all nucleons 
      with current matrix elements from 
(\ref{Definitions.nucleon.spartial}).
      Expanding the current in vector spherical harmonics one
      obtains the form 
      given in
(\ref{drateEPV}) and 
(\ref{SF-definition}):
\begin{eqnarray*} 
\frac{d\sigma^{A}_\SD}{dq^2}(v,q^2)
= \frac{S^{A}_\SD (q^2)}{v^2(2J+1)}, 
\qquad
S^A_\SD(q) 
y= \sum_{L\ \rm odd} \big( \vert\langle N \vert\vert {\cal T}^{el5}_L
(q) \vert\vert N \rangle\vert^2 + \vert\langle N \vert\vert {\cal L}^5_L
(q) \vert\vert N \rangle\vert^2\big).
\end{eqnarray*}
	The transverse electric ${\cal T}^{el5}(q)$ 
	and longitudinal ${\cal L}^5(q)$ multipole projections of the
	axial vector current operator are given 
by (see 
\cite{Engel:1992bf,Engel:1991wq,Ressell:1993qm,Ressell:1997kx}
for details):
\begin{eqnarray*}
{\cal T}^{el5}_L(q) 
        &=& \frac{1}{\sqrt{2L+1}}\sum^A_i\frac{a_0 +a_1\tau^i_3}{2}
                 \Bigl[
		-\sqrt{L}   M_{L,L+1}(q\vec{r}_i)
                +\sqrt{L+1} M_{L,L-1}(q\vec{r}_i)
                 \Bigr], \nonumber \\
 {\cal L}^5_L(q)
         &=& \frac{1}{\sqrt{2L+1}}\sum^A_i
                  \Bigl( \frac{a_0}{2} +
                     \frac{a_1 m^2_\pi \tau^i_3}{2(q^2+m_\pi^2)}
                  \Bigr) 
                  \Bigl[
                 \sqrt{L+1} M_{L,L+1}(q\vec{r}_i)
		+\sqrt{L}   M_{L,L-1}(q\vec{r}_i)
                 \Bigr], 
\end{eqnarray*} 
       where
       $ M_{L,L'}(q\vec{r}_i) =
       j_{L'}(qr_i)[Y_{L'}(\hat{r}_i)\vec{\sigma}_i]^L$, 
       $m_\pi$ is the pion mass
       and $a^{}_{0(1)}$ is the isoscalar (isovector) effective 
       spin-dependent WIMP-nucleon coupling. 
       For example, the 
       above-mentioned matrix elements one can calculate within the 
       harmonic oscillator approach on the basis of results given in    
\cite{Donnelly:1979aa}.

\bigskip 

{\small
\providecommand{\href}[2]{#2}\begingroup\raggedright\endgroup

}


\begin{thebibliography}{10}

\bibitem{Engel:1991wq}
{\em Engel J. // } Phys. Lett. B. 1991. {V.264.}
P.114--119.

\bibitem{Bottino:2003cz}
{\em Bottino A., Donato F., Fornengo N., and Scopel S. // }
\href{http://www.arXiv.org/abs/hep-ph/0307303}hep-ph/0307303.

\bibitem{Bednyakov:1994te}
{\em Bednyakov V.~A., Klapdor-Kleingrothaus H.~V., and Kovalenko S.~G. // }
  Phys. Lett. B. 1994. {V.329.} P.5--9,
\href{http://www.arXiv.org/abs/hep-ph/9401271}hep-ph/9401271.

\bibitem{Bednyakov:2000he}
{\em Bednyakov V.~A. and Klapdor-Kleingrothaus H.~V. // } Phys. Rev. D. 2001.
  {V.63.} P.095005,
\href{http://www.arXiv.org/abs/hep-ph/0011233}hep-ph/0011233.

\bibitem{Bednyakov:2002mb}
{\em Bednyakov V.~A. // } Phys. Atom. Nucl. 2003. {V.66.} P.490--493,
\href{http://www.arXiv.org/abs/hep-ph/0201046}hep-ph/0201046.

\bibitem{Bednyakov:2003wf}
{\em Bednyakov V.~A. // } Phys. Atom. Nucl. 2004. {V.67.} P.1931--1941,
\href{http://www.arXiv.org/abs/hep-ph/0310041}hep-ph/0310041.

\bibitem{Bednyakov:2004xq}
{\em Bednyakov V.~A. and Simkovic F. // } Phys. Part. Nucl. 2005. {V.36.}
  P.131--152,
\href{http://www.arXiv.org/abs/hep-ph/0406218}hep-ph/0406218.

\bibitem{Bednyakov:2005qp}
{\em Bednyakov V.~A. and Klapdor-Kleingrothaus H.~V. // }
\href{http://www.arXiv.org/abs/hep-ph/0504031}hep-ph/0504031.

\bibitem{Girard:2005pt}
{\em Girard T.~A. et al.  // } Phys. Lett. B. 2005. {V.621.} P.233--238,
\href{http://www.arXiv.org/abs/hep-ex/0505053}hep-ex/0505053.

\bibitem{Girard:2005dq}
{\em Girard T.~A. et al.  // }
\href{http://www.arXiv.org/abs/hep-ex/0504022}hep-ex/0504022.

\bibitem{Giuliani:2004uk}
{\em Giuliani F. // } Phys. Rev. Lett. 2004. {V.93.} P.161301,
\href{http://www.arXiv.org/abs/hep-ph/0404010}hep-ph/0404010.

\bibitem{Giuliani:2005bd}
{\em Giuliani F. and Girard T.~A. // } Phys. Rev. D. 2005. {V.71.} P.123503,
\href{http://www.arXiv.org/abs/hep-ph/0502232}hep-ph/0502232.

\bibitem{Savage:2004fn}
{\em Savage C., Gondolo P., and Freese K. // } Phys. Rev. D. 2004. {V.70.}
  P.123513,
\href{http://www.arXiv.org/abs/astro-ph/0408346}astro-ph/0408346.

\bibitem{Benoit:2004tt}
{\em Benoit A. et al.  // } Phys. Lett. B. 2005. {V.616.} P.25--30,
\href{http://www.arXiv.org/abs/astro-ph/0412061}astro-ph/0412061.

\bibitem{Tanimori:2003xs}
{\em Tanimori T. et al.  // } Phys. Lett. B. 2004. {V.578.} P.241--246,
\href{http://www.arXiv.org/abs/astro-ph/0310638}astro-ph/0310638.

\bibitem{Ovchinnikov:2003AA}
Ovchinnikov B.~M. and Parusov V.~V. ``The preparing of an experiment for
  search the spin-dependent interactions of wimp.'' // INR preprint 1097/2003,
  2003.

\bibitem{Moulin:2005sx}
{\em Moulin E., Mayet F., and Santos D. // } Phys. Lett. B. 2005. {V.614.}
  P.143--154,
\href{http://www.arXiv.org/abs/astro-ph/0503436}astro-ph/0503436.

\bibitem{Mayet:2002ke}
{\em Mayet F., Santos D., Bunkov Y.~M., Collin E., and Godfrin H. // } Phys.
  Lett. B. 2002. {V.538.} P.257,
\href{http://www.arXiv.org/abs/astro-ph/0201097}astro-ph/0201097.

\bibitem{Klapdor-Kleingrothaus:2005rn}
{\em Klapdor-Kleingrothaus H.~V., Krivosheina I.~V., and Tomei C. // } Phys.
  Lett. B. 2005. {V.609.} P.226--231.

\bibitem{Alner:2004cw}
{\em Alner G.~J. et al.  // } Nucl. Instrum. Meth. A. 2004. {V.535.}
P.644--655.

\bibitem{Snowden-Ifft:1999hz}
{\em Snowden-Ifft D.~P., Martoff C.~J., and Burwell J.~M. // } Phys. Rev. D. 2000. {V.61.} P.101301,
\href{http://www.arXiv.org/abs/astro-ph/9904064}astro-ph/9904064.

\bibitem{Gaitskell:1996cv}
{\em Gaitskell R.~J. et al.  // } Nucl. Instrum. Meth. A. 1996. {V.370.}
P.162--164.

\bibitem{Sekiya:2004ma}
{\em Sekiya H., Minowa M., Shimizu Y., Suganuma W., and Inoue Y. // }
\href{http://www.arXiv.org/abs/astro-ph/0405598}astro-ph/0405598.

\bibitem{Morgan:2004ys}
{\em Morgan B., Green A.~M., and Spooner N.~J.~C. // } Phys. Rev. D. 2005.
  {V.71.} P.103507,
\href{http://www.arXiv.org/abs/astro-ph/0408047}astro-ph/0408047.

\bibitem{Vergados:2000cp}
{\em Vergados J.~D. // } Part. Nucl. Lett. 2001. {V.106.} P.74--108,
\href{http://www.arXiv.org/abs/hep-ph/0010151}hep-ph/0010151.

\bibitem{Vergados:2002bb}
{\em Vergados J.~D. // } Phys. Atom. Nucl. 2003. {V.66.} P.481--489,
\href{http://www.arXiv.org/abs/hep-ph/0201014}hep-ph/0201014.

\bibitem{Jungman:1996df}
{\em Jungman G., Kamionkowski M., and Griest K. // } Phys. Rept. 1996. {V.267.}
  P.195--373,
\href{http://www.arXiv.org/abs/hep-ph/9506380}hep-ph/9506380.

\bibitem{Divari:2000dc}
{\em Divari P.~C., Kosmas T.~S., Vergados J.~D., and Skouras L.~D. // } Phys.
  Rev. C. 2000. {V.61.}
P.054612.

\bibitem{Bednyakov:1997ax}
{\em Bednyakov V.~A., Klapdor-Kleingrothaus H.~V., and Kovalenko S.~G. // }
  Phys. Rev. D. 1997. {V.55.} P.503--514,
\href{http://www.arXiv.org/abs/hep-ph/9608241}hep-ph/9608241.

\bibitem{Bernabei:2000qi}
{\em Bernabei R. et al.  // } Phys. Lett. B. 2000. {V.480.}
P.23--31.

\bibitem{Bernabei:2003za}
{\em Bernabei R. et al.  // } Riv. Nuovo Cim. 2003. {V.26.} P.1--73,
\href{http://www.arXiv.org/abs/astro-ph/0307403}astro-ph/0307403.

\bibitem{Bernabei:2003wy}
{\em Bernabei R. et al.  // }
\href{http://www.arXiv.org/abs/astro-ph/0311046}astro-ph/0311046.

\bibitem{Lewin:1996rx}
{\em Lewin J.~D. and Smith P.~F. // } Astropart. Phys. 1996. {V.6.}
P.87--112.

\bibitem{Smith:1990kw}
{\em Smith P.~F. and Lewin J.~D. // } Phys. Rept. 1990. {V.187.}
P.203.

\bibitem{Bednyakov:1999yr}
{\em Bednyakov V.~A. and Klapdor-Kleingrothaus H.~V. // } Phys. Atom. Nucl.
  1999. {V.62.}
P.966--974.

\bibitem{Bednyakov:1996yt}
{\em Bednyakov V.~A., Kovalenko S.~G., and Klapdor-Kleingrothaus H.~V. // }
  Phys. Atom. Nucl. 1996. {V.59.}
P.1718--1727.

\bibitem{Bednyakov:1997jr}
{\em Bednyakov V.~A., Kovalenko S.~G., Klapdor-Kleingrothaus H.~V., and
  Ramachers Y. // } Z. Phys. A. 1997. {V.357.} P.339--347,
\href{http://www.arXiv.org/abs/hep-ph/9606261}hep-ph/9606261.

\bibitem{Bednyakov:1994qa}
{\em Bednyakov V.~A., Klapdor-Kleingrothaus H.~V., and Kovalenko S. // } Phys.
  Rev. D. 1994. {V.50.} P.7128--7143,
\href{http://www.arXiv.org/abs/hep-ph/9401262}hep-ph/9401262.

\bibitem{Engel:1992bf}
{\em Engel J., Pittel S., and Vogel P. // } Int. J. Mod. Phys. 1992. {V.E1.}
P.1--37.

\bibitem{Ressell:1993qm}
{\em Ressell M.~T. et al.  // } Phys. Rev. D. 1993. {V.48.}
P.5519--5535.

\bibitem{Ressell:1997kx}
{\em Ressell M.~T. and Dean D.~J. // } Phys. Rev. C. 1997. {V.56.} P.535--546,
\href{http://www.arXiv.org/abs/hep-ph/9702290}hep-ph/9702290.

\bibitem{Engel:1995gw}
{\em Engel J., Ressell M.~T., Towner I.~S., and Ormand W.~E. // } Phys. Rev. C.
  1995. {V.52.} P.2216--2221,
\href{http://www.arXiv.org/abs/hep-ph/9504322}hep-ph/9504322.

\bibitem{Tovey:2000mm}
{\em Tovey D.~R., Gaitskell R.~J., Gondolo P., Ramachers Y., and Roszkowski L.
  // } Phys. Lett. B. 2000. {V.488.} P.17--26,
\href{http://www.arXiv.org/abs/hep-ph/0005041}hep-ph/0005041.

\bibitem{Vergados:1996hs}
{\em Vergados J.~D. // } J. Phys. G. 1996. {V.22.} P.253--272,
\href{http://www.arXiv.org/abs/hep-ph/9504320}hep-ph/9504320.

\bibitem{Dimitrov:1995gc}
{\em Dimitrov V., Engel J., and Pittel S. // } Phys. Rev. D. 1995. {V.51.}
  P.291--295,
\href{http://www.arXiv.org/abs/hep-ph/9408246}hep-ph/9408246.

\bibitem{Engel:1992qb}
{\em Engel J., Pittel S., Ormand E., and Vogel P. // } Phys. Lett. B. 1992.
  {V.275.} P.119--123.

\bibitem{Nikolaev:1993vw}
{\em Nikolaev M.~A. and Klapdor-Kleingrothaus H.~V. // } Z. Phys. A. 1993.
  {V.345.}
P.183--186.

\bibitem{Kosmas:1997jm}
{\em Kosmas T.~S. and Vergados J.~D. // } Phys. Rev. D. 1997. {V.55.}
  P.1752--1764,
\href{http://www.arXiv.org/abs/hep-ph/9701205}hep-ph/9701205.

\bibitem{Pacheco:1989jz}
{\em Pacheco A.~F. and Strottman D. // } Phys. Rev. D. 1989. {V.40.}
P.2131--2133.

\bibitem{Iachello:1991ut}
{\em Iachello F., Krauss L.~M., and Maino G. // } Phys. Lett. B. 1991. {V.254.}
P.220--224.

\bibitem{Ellis:1988sh}
{\em Ellis J.~R. and Flores R.~A. // } Nucl. Phys. B. 1988. {V.307.}
P.883.

\bibitem{Ellis:1991ef}
{\em Ellis J.~R. and Flores R.~A. // } Phys. Lett. B. 1991. {V.263.}
P.259--266.

\bibitem{Engel:1989ix}
{\em Engel J. and Vogel P. // } Phys. Rev. D. 1989. {V.40.}
P.3132--3135.

\bibitem{Ogawa:2000vi}
{\em Ogawa I. et al.  // } Nucl. Phys. A. 2000. {V.663.}
P.869--872.

\bibitem{Boukhira:2002qj}
{\em Boukhira N. et al.  // } Nucl. Phys. Proc. Suppl. 2002. {V.110.}
P.103--105.

\bibitem{Takeda:2003km}
{\em Takeda A. et al.  // } Phys. Lett. B. 2003. {V.572.} P.145--151,
\href{http://www.arXiv.org/abs/astro-ph/0306365}astro-ph/0306365.

\bibitem{Miuchi:2002zp}
{\em Miuchi K. et al.  // } Astropart. Phys. 2003. {V.19.} P.135--144,
\href{http://www.arXiv.org/abs/astro-ph/0204411}astro-ph/0204411.

\bibitem{Brown:1985aa}
{\em Brown B. and Wildenthal B. // } At. Data Nucl. Data Tab. 1985. {V.33.}
P.347.

\bibitem{Brown:1988aa}
{\em Brown B. and Wildenthal B. // } Ann. Rev. Nucl. Part. Sci. 1988. {V.38.}
P.29.

\bibitem{Bernabei:2003ky}
{\em Bernabei R. et al.  // } Nucl. Phys. A. 2003. {V.719.}
P.257--265.

\bibitem{Ahmed:2003su}
{\em Ahmed B. et al.  // } Astropart. Phys. 2003. {V.19.} P.691--702,
\href{http://www.arXiv.org/abs/hep-ex/0301039}hep-ex/0301039.

\bibitem{Cebrian:2002vd}
{\em Cebrian S. et al.  // } Nucl. Phys. Proc. Suppl. 2003. {V.114.}
  P.111--115,
\href{http://www.arXiv.org/abs/hep-ex/0211050}hep-ex/0211050.

\bibitem{Yoshida:2000df}
{\em Yoshida S. et al.  // } Nucl. Phys. Proc. Suppl. 2000. {V.87.}
P.58--60.

\bibitem{Angloher:2003cg}
{\em Angloher G. et al.  // } Phys. Atom. Nucl. 2003. {V.66.}
P.494--496.

\bibitem{Resler:1988aa}
{\em Resler D. and Grimes S. // } Computers in Phys. 1988. {V.2 (3).}
P.65--66.

\bibitem{Klapdor-Kleingrothaus:2002pg}
{\em Klapdor-Kleingrothaus H.~V. et al.  // } Astropart. Phys. 2003.
  {V.18.} P.525--530,
\href{http://www.arXiv.org/abs/hep-ph/0206151}hep-ph/0206151.

\bibitem{Petrovich:1969aa}
{\em Petrovich F., McManus H., Madsen V., and Atkinson J. // } Phys. Rev. Lett.
  1969. {V.38.}
P.895.

\bibitem{Dimitrov:1994aa}
{\em Dimitrov V. // } Phys. Rev. C. 1994. {V.50.}
P.2893--2899.

\bibitem{Nikolaev:1993dd}
{\em Nikolaev M.~A. and Klapdor-Kleingrothaus H.~V. // } Z. Phys. A. 1993.
  {V.345.}
P.373--376.

\bibitem{Hjorth-Jensen:1995ap}
{\em Hjorth-Jensen M., Kuo T.~T.~S., and Osnes E. // } Phys. Rept. 1995.
  {V.261.}
P.125--270.

\bibitem{Stoks:1994wp}
{\em Stoks V.~G.~J., Klomp R.~A.~M., Terheggen C.~P.~F., and de Swart~J.~J. //
  } Phys. Rev. C. 1994. {V.49.} P.2950--2962,
\href{http://www.arXiv.org/abs/nucl-th/9406039}nucl-th/9406039.

\bibitem{Hart:2002qh}
{\em Hart S.~P. // } Nucl. Phys. Proc. Suppl. 2002. {V.110.}
P.91--93.

\bibitem{Luscher:2003er}
{\em Luscher R. // }
\href{http://www.arXiv.org/abs/astro-ph/0305310}astro-ph/0305310.

\bibitem{Bernabei:2002qg}
{\em Bernabei R. et al.  // } Nucl. Phys. Proc. Suppl. 2002. {V.110.}
P.88--90.

\bibitem{Vergados:1996jp}
{\em Vergados J.~D. and Kosmas T.~S. // }
\href{http://www.arXiv.org/abs/hep-ph/9701204}hep-ph/9701204.

\bibitem{Vergados:1996jn}
{\em Vergados J.~D. and Kosmas T.~S. // }
\href{http://www.arXiv.org/abs/hep-ph/9701206}hep-ph/9701206.

\bibitem{Kosmas:1995md}
{\em Kosmas T.~S. and Vergados J.~D. // }
\href{http://www.arXiv.org/abs/nucl-th/9509026}nucl-th/9509026.

\bibitem{Donnelly:1979aa}
{\em Donnelly T. and Haxton W. // } At. Data Nucl. Dat. Tab. 1979. {V.23.}
P.103--176.

\end{thebibliography}
\end{document}